\documentclass[12pt]{article}

\usepackage[utf8]{inputenc}
\usepackage[T1]{fontenc}
\usepackage{amsmath,amssymb,amsfonts,amsthm}
\newtheorem{proposition}{Proposition}
\newtheorem{lemma}{Lemma}
\newtheorem{corollary}{Corollary}
\theoremstyle{remark}
\newtheorem{remark}{Remark}
\usepackage{algorithm}
\usepackage{algpseudocode}
\usepackage{booktabs}
\usepackage{graphicx}
\usepackage{natbib}
\usepackage{hyperref}
\usepackage{geometry}
\usepackage{bm}
\usepackage{mathtools}
\usepackage{tikz}
\usetikzlibrary{arrows.meta,positioning,shapes.geometric,fit,backgrounds}
\newcommand{\authororcid}[1]{%
  \href{https://orcid.org/#1}{\texttt{\footnotesize ORCID: #1}}}
\newcommand{\mscclass}[1]{%
  \par\medskip\noindent
  \textbf{Mathematics Subject Classification (2020):} #1}

\newcommand{\coefchart}[1]{%
  \draw[gray!60,-{Stealth[length=1.6mm]}] (-0.15,0) -- (2.35,0);
  \draw[gray!60] (0,-0.55) -- (0,0.8);
  \foreach \i/\v in {#1}{%
    \pgfmathsetmacro{\xx}{0.18+\i*0.36}%
    \ifdim\v cm<0cm
      \fill[red!55] (\xx-0.09,0) rectangle (\xx+0.09,\v);
    \else
      \fill[blue!60] (\xx-0.09,0) rectangle (\xx+0.09,\v);
    \fi
  }%
}

\newcommand{\R}{\mathbb{R}}
\newcommand{\bX}{\bm{X}}
\newcommand{\bY}{\bm{Y}}
\newcommand{\bT}{\bm{T}}
\newcommand{\bU}{\bm{U}}
\newcommand{\bW}{\bm{W}}
\newcommand{\bP}{\bm{P}}
\newcommand{\bQ}{\bm{Q}}
\newcommand{\bB}{\bm{B}}
\newcommand{\bV}{\bm{V}}
\newcommand{\ba}{\bm{a}}
\newcommand{\bc}{\bm{c}}
\newcommand{\indep}{\mathrel{\perp\!\!\!\perp}}
\newcommand{\bw}{\bm{w}}
\newcommand{\bt}{\bm{t}}
\newcommand{\bu}{\bm{u}}
\newcommand{\bv}{\bm{v}}
\newcommand{\bp}{\bm{p}}
\newcommand{\bq}{\bm{q}}
\newcommand{\bxi}{\bm{x}_i}
\newcommand{\byi}{\bm{y}_i}
\newcommand{\Var}{\operatorname{Var}}
\newcommand{\E}{\operatorname{\mathbb{E}}}
\newcommand{\tbtree}{\textsf{tbtree}}
\newcommand{\twoblock}{\textsf{twoblock}}

\newcommand{\rev}[1]{#1}
\newcommand{\revdel}[1]{}

\title{Twoblock clustering trees with coskewness-based dimension
       reduction: recovering piecewise multivariate linear regimes%
  \thanks{The Supplementary Material referred to throughout the text is
  appended to this preprint as Sections~S1--S9, following Appendix~A.}}

\author{Sven Serneels$^{1,2}$ \\[6pt]
\small $^1$ Snow Stallion AI, Cheyenne, Wyoming, USA \\
\small $^2$ Department of Mathematics, University of Antwerp, Belgium%
\footnote{Corresponding author, e-mail: \texttt{sven.serneels@uantwerpen.be}}%
\\[2pt]\authororcid{0000-0003-1642-189X}
}
\date{}

\begin{document}
\maketitle

% =====================================================================
\begin{abstract}
% =====================================================================
The twoblock clustering tree (\tbtree) is introduced as a highly interpretable regression tree for multivariate responses. Twoblock trees are deterministic decision trees that have local multivariate linear models as their leaves and use dense or sparse twoblock dimension reduction as local leaf models and in the impurity.  The resulting models are both computationally efficient and can be highly interpretable.  \rev{The estimator's primary aim is an interpretable, regime-aligned piecewise-linear description of the data, with predictive competitiveness retained as a constraint through a split/leaf decoupling.}  Beyond proposing the decision tree estimator itself,  this paper also introduces an estimator for the twoblock dimension reduced space based on maximizing coskewness, which facilitates identification of non-normal clusters in the data. The tree inherently produces a set of local linear models and is therefore apt to recover \rev{piecewise} linear regimes, which is illustrated in a simulation. However, two \rev{real-world} data examples illustrate that twoblock trees are also capable of modeling more complexly nonlinear dependencies and can perform on par with \rev{black-box} modeling techniques, such as random forests. At each point, both the twoblock models that generate the splits, as well as the ones in the leaves, can be inspected and interpreted.


\mscclass{62H30, 62H25, 62J05, 62-08}
\end{abstract}

% =====================================================================
\section{Introduction}\label{sec:intro}
% =====================================================================

% --- 1.1 Non-linearity calls for flexible estimators ---

In many industrial and scientific applications the mechanism that
links a set of predictors to the quantities one wishes to predict is
genuinely non-linear, yet it can reasonably be approximated by a set of local linear regimes. Manufacturing processes are often operated in distinct modes, for instance to produce different grades. In materials science, properties may depend on composition class. Likewise, biological responses may stratify by genotype. In all such cases, the response surface is shaped by discrete or threshold mechanisms on top of smooth local trends, and a single global linear model is forced to average across the underlying
structure. To accurately predict such responses, a global model is required that can approximate a structure with many linear regimes closely. Sundry models exist that can approximate nonlinear responses very well, such as random forests or neural networks.

% --- 1.2 But we also want to understand the model ---

Predictive accuracy, however, is rarely the only requirement. In the
same applications, practitioners typically also need to understand
\emph{how} a model reaches its predictions -- to validate it against
domain knowledge, to diagnose failures, and, increasingly, to satisfy
regulatory demands for transparency. What is needed are methods that
are simultaneously accurate and \emph{highly interpretable}. The
flexible regressors that dominate the accuracy benchmarks, such as random
forests \citep{Breiman2001RF} and multilayer perceptrons \citep{Goodfellow2016DL}, sit at the opposite end of this trade-off:
a $200$-tree forest contains hundreds of thousands of split parameters
and a feed-forward network thousands of trainable weights, so that
such models become very hard to inspect or interpret.

% --- 1.3 Trees with models in the leaves ---

The decision-tree family has presented itself as a strong candidate for a long time to produce both highly interpretable {\em and} highly accurate fits and predictions. Because such models arrive at decisions through a short sequence of explicit conditions, they are easy to explain or inspect. The
classical regression tree of \citet{Breiman1984} fits a constant in
every leaf and remains one of the most readily interpretable
predictive models available, but it tends either to over-fit or to
require very deep trees with many leaves to reach competitive accuracy.  A deep tree, however, is no longer interpretable in any practical sense.
Random forests resolve the accuracy problem by averaging over many such
trees, but abandon interpretability entirely. For these reasons a
substantial literature has proposed decision trees that carry a
\emph{regression model in each leaf} rather than a constant, so that
fewer, shallower splits suffice.

The oldest family of this kind is due to Quinlan: the M5 model tree
\citep{Quinlan1992M5} places a linear regression model in every leaf,
and its rule-based descendant Cubist \citep{Quinlan1993Cubist} recasts
the tree as a collection of rules, each equipped with its own linear
model, and remains in routine use today. In practice, however, Cubist
too reaches competitive accuracy only when several such rule-based
models are combined into \emph{committees} -- a boosting-style ensemble
in which each successive member corrects the residuals of its
predecessors. A single model tree is often too coarse: its
axis-aligned splits on individual predictors cannot capture regime
boundaries that run obliquely through the predictor space, so accuracy
is recovered by averaging many imperfect trees instead. The price is
the familiar one: a committee of overlapping rule sets can no longer
be read or validated as one transparent model, so the very property
that motivated the model-in-leaf construction is forfeited again.
This has spurred the search for methods that attain accuracy within a
\emph{single} tree. \citet{ErikssonTryggWold2009} bridged
dimension reduction and trees in exactly this spirit with their
\emph{PLS-Trees}, which fit a partial least squares (PLS) regression at each
node and split along the first predictor score, yielding leaves whose
loading vectors expose interpretable local sub-models. A parallel line
of work seeks globally optimal trees by mixed-integer optimization
\citep{BertsimasDunn2017,BertsimasDunnYang2021}; these are elegant but
computationally prohibitive at realistic scale -- \citet{BertsimasDunn2017}
report per-tree time limits of thirty minutes to two hours, never test
trees deeper than four levels, and target a single response rather than
the multivariate case.

% --- 1.4 The twoblock base and the regime-identification observation ---

Novel in this paper is to initiate tree construction from the recently proposed {\em twoblock}
method \citep{Cook2023}, which returns, for a
predictor block $\bX \in \R^{n\times p}$ and a response block
$\bY \in \R^{n\times q}$, a reduced central subspace of each block with
respect to the other. It is a genuinely multivariate joint reduction with optional 
embedded
per-block variable selection if the sparse variant \rev{of} \citet{Serneels2025sparse} is selected. When the data contain several operating
regimes a single global twoblock model fits poorly; but, owing to its
construction, the reduced score spaces it produces are an excellent
device for \emph{identifying} those regimes as distinct subspaces. This
is the observation exploited here: rather than fight the global misfit,
the twoblock scores are used to partition the data and place a fresh,
locally faithful twoblock model in each part.

% --- 1.5 Coskewness: the key methodological contribution ---

The published twoblock methods choose their latent directions to
maximize the \emph{covariance} of the predictor and response scores.
This is the natural objective when the two blocks are jointly Gaussian,
but it may be the wrong lens for the regime-identification task.
A node that straddles two regimes contains a \emph{mixture} of two
local linear laws, and a mixture of Gaussians is non-Gaussian: its
skewness and higher moments carry the signature of the components.
Concretely, the direction that best separates two regimes is often one
along which the joint predictor--response distribution is strongly
\emph{skewed} while its covariance with the response is weak or even
zero. The covariance objective is blind to such a direction by
construction; a third-moment objective is not. This paper therefore
introduces, nested into its central methodological contribution, a novel
twoblock decomposition whose latent directions maximize the
\emph{coskewness} of the scores rather than their covariance, or a convex combination of the two criteria that trades regime
sensitivity against predictive sharpness. Both are computed by a
matrix-free, block-symmetric higher-order power method
\citep{DeLathauwer2000hopm} that never materializes a third-moment
tensor, so the cost per latent direction stays linear in the data size.

% --- 1.6 Contributions ---

Putting these pieces together, the twoblock clustering tree
(\tbtree) is proposed: a potentially shallow, interpretable, deterministic top-down clustering scheme that fits a twoblock model at each node, splits along the first predictor
score, and is multivariate by construction. Its specific contributions
are the following.  At first, twoblock trees \rev{proposed} here (\tbtree) are the first tree-based regression method to place a local, genuinely multivariate twoblock (dense or sparse) dimension reduction model in every leaf. They generalize the PLS-Trees of \citet{ErikssonTryggWold2009} from partial least squares to the joint predictor--response reduction of \citet{Cook2023}, so both the split direction and the leaf prediction are informed by the multivariate response block.  Secondly,  to obtain a split impurity more efficient at detecting regimes, a novel twoblock decomposition is introduced based on \emph{coskewness}: its latent directions maximize a third-order comoment of the predictor and response scores, or a convex combination with the classical covariance objective. \revdel{This dimension reduction based on thrid order comoments is novel by itself and has not been published elsewhere.}\rev{To the best of the author's knowledge, this dimension reduction based on third-order comoments is novel in its own right.}

\rev{Since the method touches both regime detection and prediction,
it is worth stating explicitly what the primary statistical goal of
\tbtree\ is. The primary goal is an interpretable,
\emph{regime-aligned} piecewise multivariate linear description of
the data: a shallow partition whose cells track the latent regimes
and whose local models can be read. The regime claim is deliberately a \emph{relative} one:  the third-moment split produces partitions substantially more aligned with the true regime structure than
covariance-based splitting does.  It is not a claim of exact regime identification, which greedy binary cuts on a single score, combined
with the sampling variability of third moments, cannot deliver in
general. Prediction enters as a constraint rather than as the
objective and the tree's decoupled design (Section~\ref{sec:tbtree-induction}) is built precisely so
that pursuing the first goal does not surrender the second: a
third-moment \emph{split} objective buys regime-aligned routing,
while covariance \emph{leaf} models keep prediction sharp. The two empirical chapters divide along the same line. The simulation of
Section~\ref{sec:tbtree-sim}, where ground-truth regime labels exist, measures the alignment gain (and its limits) directly, while quantifying the trade-off that a coupled third-moment tree
incurs; the real-data benchmarks of Section~\ref{sec:examples},
where no labels exist, verify the constraint of having predictive parity
with black-box references, while the structural payoff appears there as the interpretability of the fitted regime models.}

The remainder of the paper is organized as follows.
Section~\ref{sec:twoblock-recap} recaps the twoblock estimator.
Section~\ref{sec:coskew} develops the coskewness decomposition and its
higher-order power method. Section~\ref{sec:coskew-sim} reports a
controlled simulation for the coskewness objective in isolation.
Section~\ref{sec:tbtree} states the \tbtree\ algorithm, its impurity
criterion and its cross-validation pruning gate.
Section~\ref{sec:tbtree-sim} reports the four-regime simulation study.
Section~\ref{sec:examples} applies the method to the energy-efficiency
and gas-turbine benchmarks. Section~\ref{sec:conclusions}
concludes and Section~\ref{sec:software} points to the implementation.

% =====================================================================
\section{The twoblock estimator}\label{sec:twoblock-recap}
% =====================================================================

Consider $n$ rows of two blocks of variables, a predictor block
$\bX \in \R^{n\times p}$ and a response block $\bY \in \R^{n\times q}$,
with $\bxi$ and $\byi$ the $i$th rows; both blocks are centred and
scaled before the decomposition. The twoblock estimator
\citep{Cook2023} performs a \emph{simultaneous sufficient dimension
reduction} of the two blocks: it seeks score matrices $\bT = \bX\bW$
and $\bU = \bY\bV$, of possibly different ranks $h$ and $g$, such that
\begin{equation}
\bY \indep \bX \mid \bT
\qquad\text{and}\qquad
\bX \indep \bY \mid \bU,
\label{eq:msdr}
\end{equation}
i.e.\ each block is reduced to an estimate of a central subspace that
carries all the information that block holds about the \emph{other}
block. In sample, with the independence conditions relaxed to zero
covariance, \citet{Cook2023} estimate the two subspaces by a pair of
PLS-type deflation loops, one per block. Let $\bX_0 = \bX$ and
$\bY_0 = \bY$; the weight vectors maximize the covariance between a
candidate score in the (deflated) own block and the other block in
full,
\begin{subequations}\label{eq:cov-objective}
\begin{align}
\bw_i &= \operatorname*{argmax}_{\|\bc\|=1}\;
          \bc^\top \bX_{i-1}^\top \bY \bY^\top \bX_{i-1}\, \bc,
& \bt_i &= \bX_{i-1}\bw_i,
\label{eq:cov-objective-x}\\
\bv_j &= \operatorname*{argmax}_{\|\ba\|=1}\;
          \ba^\top \bY_{j-1}^\top \bX \bX^\top \bY_{j-1}\, \ba,
& \bu_j &= \bY_{j-1}\bv_j,
\label{eq:cov-objective-y}
\end{align}
\end{subequations}
each loop followed by its own rank-one deflation
$\bX_i = \bX_{i-1} - \bt_i\bp_i^\top$ and
$\bY_j = \bY_{j-1} - \bu_j\bq_j^\top$, with loadings
$\bp_i = \bX_{i-1}^\top\bt_i/\bt_i^\top\bt_i$ and
$\bq_j = \bY_{j-1}^\top\bu_j/\bu_j^\top\bu_j$, which keeps successive
scores within each block uncorrelated. For the leading component the
two criteria are solved jointly by the dominant singular vector pair of
$\bX^\top\bY/n$, i.e.\ $(\bw_1,\bv_1)$ maximize the score covariance
$\E[t\,u]$, which is equivalent to SIMPLS \citep{deJong1993}. The difference from \rev{multi-response PLS (PLS2)} or
SIMPLS resides beyond the first component: there, the $\bY$-side scores and loadings
are a computational by-product of a single $\bX$-driven deflation,
whereas twoblock deflates each block separately, so that $\bT$
\emph{and} $\bU$ are both genuine central-subspace estimates and the
two blocks may retain different numbers of components. The two
reductions recombine into a single coefficient matrix
\begin{equation}
\bB \;=\;
\bW\left(\bW^\top\bX^\top\bX\bW\right)^{-1}\bW^\top\bX^\top\bY\,\bV\bV^\top
\;\in\; \R^{p\times q},
\qquad
\hat{\bY} = \bX\bB,
\label{eq:twoblock-B}
\end{equation}
the regression of $\bY$ on the $\bX$-score space, projected onto the
reduced $\bY$-subspace.

Two features matter downstream. First, the
reduction is genuinely multivariate: a single joint model relates all
$p$ predictors to all $q$ responses through a shared low-dimensional
score space. Second, \citet{Serneels2025sparse} has extended the extraction
with soft-thresholding on $\bw$ and $\bv$, governed by penalties
$\eta_x,\eta_y \in [0,1)$.  This produces a \emph{sparse}, variable-selecting
variant, which was \rev{the first of its kind} to produce dimension reduction in which both the number of components {\em and} the sparsity can be tuned individually in each block.  It is noted that a huge literature exists on sufficient variable selection (see e.g. \citet{Menvouta2022sparse} for options based on martingale difference divergence \citep{ShaoZhang2014mdd}), but \rev{none of these methods} introduce sparsity for each block independently.   Further extensions into casewise and cellwise robust estimators are given in
\citet{Serneels2025rtb,Serneels2026crtb}. The \tbtree\ estimator is agnostic to the
choice of base estimator, and its leaves may therefore carry dense or
sparse local models. While not \rev{pursued} in this paper, leaves could even be casewise or cellwise robust twoblock models.

The single most important property for what follows is that
\eqref{eq:cov-objective} is a \emph{second-moment} objective: every
quantity in the estimator derives from cross-covariances alone. When
the data within a node are homogeneous -- a single local linear law
with Gaussian scatter -- the covariance criterion is the right one.
When the node instead contains a mixture of regimes, the criterion is
no longer sufficient. For that reason, a criterion based on \rev{third-order comoments} is adopted in what follows.

% =====================================================================
\section{A coskewness-based twoblock decomposition}\label{sec:coskew}
% =====================================================================

\subsection{Why a third-moment objective}\label{sec:coskew-why}

Suppose a node contains two regimes, each a local linear map from
$\bX$ to $\bY$ with Gaussian scatter, in proportions $\pi$ and
$1-\pi$. Along any direction the marginal score distribution is then a
two-component Gaussian \emph{mixture}, which is non-Gaussian whenever
the component means differ. The signature of that non-Gaussianity is carried by the higher-order moments: a two-component mixture is, in
general, both skewed and heavy-tailed relative to a single Gaussian, and the skewness in particular encodes the \emph{asymmetry} between the
two components. The direction that best separates the regimes is therefore often a direction of large joint \emph{skewness}, and it need
not coincide with -- and may be orthogonal to -- the direction of large
joint covariance. A covariance-based objective such as
\eqref{eq:cov-objective} cannot, by construction, prefer such a direction; a third-moment objective can. This is the motivation for a
twoblock decomposition driven by \emph{coskewness}, the natural
third-order cross-moment between the predictor and response scores.
\rev{Section~\ref{sec:coskew-theory} makes this intuition precise: it
characterizes the population maximizers of the resulting criteria under an explicit mixture model, including the mixture proportions for which the third-moment signal is strong, weak, or, in the
symmetric case,  or absent altogether.  It then states the sampling conditions under which the empirical criterion inherits these properties.}

At this point, it is noted that dimension reduction based on \rev{higher-order comoments} has been explored \rev{before.} For instance, a \rev{kurtosis-based} one-block dimension reduction was illustrated to better reveal \rev{clusters} in analytical chemical data \citep{Driscoll2020sppa}. Likewise, a criterion that \rev{accommodates higher-order comoments} was adopted to calculate generalized betas for financial portfolios \citep{Serneels2019betas}. However, in both those approaches, a computationally involved projection pursuit algorithm was used.  This article is first to extend the use of \rev{higher-order comoments} into the twoblock simultaneous dimension reduction framework and it is also first to do so based on a computationally very efficient matrix-free power method, as will \rev{be} explained in Subsection \ref{sec:coskew-hopm}.

\subsection{Objectives}\label{sec:coskew-obj}

Write $\bt = \bX\bw$ and $\bu = \bY\bv$ for a generic pair of score
vectors. Beside the covariance $\E[t\,u]$ that drives
\eqref{eq:cov-objective}, the two third-order cross-moments of the pair
are $\E[t^2 u]$ and $\E[t\,u^2]$. The coskewness decomposition retains
the architecture of Section~\ref{sec:twoblock-recap}: two deflation
loops, each pairing its own deflated block with the other block in full. However, the \rev{second-order} criterion is supplanted by a criterion that involves the \rev{third-order comoment}.
With $\bX_0 = \bX$ and $\bY_0 = \bY$ as before, the $i$th $\bX$-side
weight solves
\begin{equation}
\bw_i \;=\; \operatorname*{argmax}_{\|\bw\|=\|\bv\|=1}\;
\E[\,t^2 u\,],
\qquad
t = \bX_{i-1}\bw,\quad u = \bY\bv,
\label{eq:coskew-objective}
\end{equation}
where $\E[t^2 u] = \tfrac{1}{n}\sum_{k=1}^n t_k^2\,u_k$ is the
empirical \emph{coskewness} of the score pair. The auxiliary response
weight $\bv$ is part of the maximization but only $\bw_i$ is retained
-- exactly as the covariance criterion \eqref{eq:cov-objective-x}
implicitly maximizes over a response direction, which in that case can be
eliminated analytically. The $j$th $\bY$-side weight $\bv_j$
symmetrically maximizes $\E[t\,u^2]$ with $t = \bX\bw$ and
$u = \bY_{j-1}\bv$. Scores, loadings and the per-block rank-one
deflations are exactly those of Section~\ref{sec:twoblock-recap}, so
the two loops may again run to different ranks $h$ and $g$, and the
resulting $(\bW,\bV)$ recombine into regression coefficients through
\eqref{eq:twoblock-B} unchanged: the coskewness criterion is a drop-in
replacement for the covariance criterion inside the twoblock
estimator. Because coskewness mixes third-order moments across
variables, it is not scale invariant\revdel{; one has to take care that both blocks are always autoscaled prior to calculating results, so that the objective becomes a
moment of standardized scores}\rev{; a dedicated paragraph below
details which normalizations \eqref{eq:coskew-objective}
does and does not entail, and introduces a fully scale-invariant
variant of the objective}.

Third moments are, however, estimated with more sampling variability
than second moments, and on data that are close to a single linear law
the covariance objective remains preferable. A \emph{combined}
objective interpolates between the two within each loop,
\begin{equation}
\max_{\|\bw\|=\|\bv\|=1}\;
(1-\gamma)\,\E[t\,u] + \gamma\,\E[t^2 u],
\qquad \gamma \in [0,1],
\label{eq:combined-objective}
\end{equation}
on the $\bX$-side, and symmetrically with $\E[t\,u^2]$ on the
$\bY$-side, with $\gamma=0$ recovering the covariance decomposition of
\citet{Cook2023} exactly and $\gamma=1$ the pure coskewness objective.
The weight $\gamma$ is a single interpretable knob trading regime
sensitivity (large $\gamma$) against predictive sharpness on smooth
data (small $\gamma$).
\rev{In practice $\gamma$ is selected by $K$-fold cross-validation
over a \emph{coarse} grid such as
$\gamma \in \{0, 0.25, 0.5, 0.75, 1\}$, with $\gamma=0.5$ the
default when no tuning budget is available. A coarse grid is not a
compromise but the appropriate resolution: Remark~\ref{rem:gamma} shows that the maximizer of
\eqref{eq:combined-objective} is piecewise (nearly) constant in $\gamma$, jumping between the covariance and coskewness basins at a
single crossing weight, so downstream performance is plateau-shaped in $\gamma$ and a fine grid buys nothing. The Supplementary Material documents these plateaus on all three examples of this
paper. Since the implementation is a \texttt{scikit-learn} compatible estimator, the search is a standard grid search. Note also that in the raw combined objective the two terms are moments of different
order, so the value of $\gamma$ at which the basins cross depends on the scale of the scores; under the score-standardized variant introduced below both terms are dimensionless, which makes chosen $\gamma$ values comparable across datasets and across tree nodes.}

\rev{\paragraph{Scaling and standardization.}
Two distinct scale effects must be kept apart in the third-order
objectives. The first concerns the \emph{columns} of the blocks: a variable's measurement unit enters the third-moment sums cubically, so both blocks are centred and autoscaled before the decomposition:
within \tbtree, per node on that node's own rows (Section~\ref{sec:tbtree-induction}). That makes
\eqref{eq:coskew-objective} a moment of \emph{standardized variables}. The second concerns the \emph{scores}: autoscaled columns combined ith unit-norm weights do not make the latent scores unit variance, since $\Var(t) = \bw^\top \bm{R}_X \bw$ with $\bm{R}_X$ the correlation matrix of the block, which equals one for every unit-norm $\bw$ only when the columns are uncorrelated. The raw criterion
\eqref{eq:coskew-objective} therefore rewards a direction both for its standardized skewness and for its score variance: the same design choice that separates PLS-type from CCA-type objectives at second
order, where the covariance $\E[tu]$ likewise blends correlation with score scale. Proposition~\ref{prop:coskew-ident} below shows that this is no defect for the intended use: under the mixture model of
Section~\ref{sec:coskew-theory} the population maximizer of the raw criterion is exactly the regime-separating mean-shift direction, irrespective of the within-regime covariance structure of either
block. Nevertheless, a fully scale-invariant form of the objective is well defined and is provided as an option, by standardizing the scores
inside the criterion:
\begin{equation}
\bw_i \;=\; \operatorname*{argmax}_{\bw,\,\bv}\;
\frac{\E[\,t^2 u\,]}{\Var(t)\,\sqrt{\Var(u)}},
\qquad t = \bX_{i-1}\bw,\quad u = \bY\bv,
\label{eq:std-objective}
\end{equation}
equivalently the raw criterion maximized under the constraints $\Var(t)=\Var(u)=1$ in place of $\|\bw\|=\|\bv\|=1$. In the combined objective \eqref{eq:combined-objective} the covariance term is then
replaced by the score correlation, so that both terms -- and hence the
interpolation weight $\gamma$ -- become scale-free. This \emph{standardized} variant is invariant under invertible linear transformations of either block; its population maximizer under the
mixture model is the Fisher discriminant direction of the regime shift (Corollary~\ref{cor:coskew-std}), and
Section~\ref{sec:coskew-hopm} describes how it is solved by the same matrix-free machinery on whitened blocks. Both variants are compared in Section~\ref{sec:coskew-sim} and, across mixture proportions and
sample sizes, in the supplementary material.}

Although the decomposition is defined for arbitrary ranks, the tree
induction of Section~\ref{sec:tbtree} only ever consumes its
\emph{leading} component: the split at a node is searched along the
first $\bX$-score $t_1 = \bX\bw_1$, so tree construction solves
\eqref{eq:coskew-objective} or \eqref{eq:combined-objective} in the
single-component case $i=1$, for which no deflation is required.
Multi-component coskewness fits arise only when the leaf models
themselves are requested under a third-moment objective, which is the coupled
variant discussed in Section~\ref{sec:tbtree-induction}.

\rev{\subsection{Population behaviour under a mixture of
regimes}\label{sec:coskew-theory}

This subsection makes the motivation of Section~\ref{sec:coskew-why} precise. The data-generating model is the one the tree is designed to detect: the rows of a node are drawn from a mixture of $K$ regimes.
Conditional on regime $k$, the predictor and response vectors have means $\bm{\mu}^x_k$ and $\bm{\mu}^y_k$, covariances $\bm{\Sigma}^x_k$ and $\bm{\Sigma}^y_k$, cross-covariance
$\bm{\Sigma}^{xy}_k$, and vanishing third central moments.  It is noted that within-regime Gaussianity, as in the simulations of Sections~\ref{sec:coskew-sim} and~\ref{sec:tbtree-sim}, suffices, but
any elliptical within-regime law qualifies. The blocks are centred, so the mixture mean is zero. For a direction pair $(\bw,\bv)$ write $t=\bw^\top\bm{x}$ and $u=\bv^\top\bm{y}$, and let $a_k = \bw^\top\bm{\mu}^x_k$ and $b_k = \bv^\top\bm{\mu}^y_k$ denote the projected regime means.

\begin{lemma}[moment decomposition]\label{lem:coskew-decomp}
Under the mixture model with mixing proportions
$\pi_1,\dots,\pi_K$,
\begin{equation}
\E[\,t^2u\,] \;=\; \sum_{k=1}^{K}\pi_k\Bigl(
b_k\,\bw^\top\bm{\Sigma}^x_k\bw
\;+\;2a_k\,\bw^\top\bm{\Sigma}^{xy}_k\bv\Bigr)
\;+\;\sum_{k=1}^{K}\pi_k\,a_k^2\,b_k .
\label{eq:coskew-decomp}
\end{equation}
\end{lemma}

The first sum couples the projected mean shifts to the within-regime covariance structure; the second is a pure between-regime term. For two regimes with a common within-regime covariance structure, the
first sum cancels and the criterion becomes remarkably simple.

\begin{proposition}[population identification]\label{prop:coskew-ident}
Let $K=2$ with mixing proportions $\pi$ and $1-\pi$ and common
within-regime covariances,
$\bm{\Sigma}^x_1=\bm{\Sigma}^x_2=\bm{\Sigma}^x_{\mathrm{w}}$,
$\bm{\Sigma}^y_1=\bm{\Sigma}^y_2$,
$\bm{\Sigma}^{xy}_1=\bm{\Sigma}^{xy}_2=\bm{\Sigma}^{xy}_{\mathrm{w}}$,
and write $\bm{\delta}_x = \bm{\mu}^x_2-\bm{\mu}^x_1$ and
$\bm{\delta}_y = \bm{\mu}^y_2-\bm{\mu}^y_1$ for the regime mean
shifts. Then
\begin{equation}
\E[\,t^2u\,]
\;=\;\pi(1-\pi)(2\pi-1)\,
(\bw^\top\bm{\delta}_x)^2\,(\bv^\top\bm{\delta}_y).
\label{eq:coskew-ident}
\end{equation}
Consequently, whenever $\pi\neq\tfrac12$, $\bm{\delta}_x\neq\bm{0}$
and $\bm{\delta}_y\neq\bm{0}$, the maximizers of
\eqref{eq:coskew-ident} over unit-norm $(\bw,\bv)$ are
$\bw^\ast=\pm\,\bm{\delta}_x/\|\bm{\delta}_x\|$ together with
$\bv^\ast=s\,\bm{\delta}_y/\|\bm{\delta}_y\|$,
$s=\operatorname{sign}(2\pi-1)$, with maximal value
$\pi(1-\pi)\,|2\pi-1|\,\|\bm{\delta}_x\|^2\,\|\bm{\delta}_y\|$; and
the criterion vanishes identically in $(\bw,\bv)$ if and only if
$\pi=\tfrac12$ or one of the shifts is zero.
\end{proposition}

Proofs are collected in Appendix~\ref{app:proofs}. Three features of Proposition~\ref{prop:coskew-ident} answer, in turn, how the raw criterion depends on the covariance structure of the two blocks, what
happens for symmetric mixtures, and how the mixing proportion modulates the signal.

First, the within-regime covariances drop out of \eqref{eq:coskew-ident} entirely: at the population level the raw coskewness criterion is maximized \emph{exactly} at the mean-shift
direction pair, whatever the covariance structure within the regimes and however the two blocks are correlated there. The covariance objective enjoys no such property: under the same model,
$\E[tu]=\bw^\top\bigl(\bm{\Sigma}^{xy}_{\mathrm{w}}
+\pi(1-\pi)\,\bm{\delta}_x\bm{\delta}_y^\top\bigr)\bv$, so its maximizer blends the within-regime regression structure with the regime shift and recovers the shift direction only when the
between-regime term happens to dominate. This is the formal counterpart of the decoy phenomenon of Section~\ref{sec:coskew-sim}. When the within-regime covariances do differ across regimes, the cancelled terms reappear as the first sum of \eqref{eq:coskew-decomp} and perturb the maximizer continuously in the magnitude of that heterogeneity.

Second, at $\pi=\tfrac12$ the factor $2\pi-1$ annihilates the criterion: a balanced two-component mixture carries no third-moment signal in any direction, at any sample size: its bimodality is a
fourth-moment phenomenon. This is an intrinsic limitation of any skewness-based criterion, not of its estimator. It is the principal reason the combined objective \eqref{eq:combined-objective} retains a
covariance term ($\gamma<1$) as its recommended form, and it marks the natural boundary of the present method: near-balanced regime structures call for a cokurtosis extension, taken up in the outlook of Section~\ref{sec:conclusions}.

Third, the prefactor $\pi(1-\pi)\,|2\pi-1|$, maximal at $\pi = \tfrac12 \pm 1/(2\sqrt{3}) \approx 0.21,\ 0.79$, quantifies how the strength of the population signal varies with the mixture balance and enters the finite-sample behaviour directly, as made explicit in
Remark~\ref{rem:samplesize} below.

\begin{corollary}[standardized criterion]\label{cor:coskew-std}
Under the assumptions of Proposition~\ref{prop:coskew-ident}, the maximizer of the standardized criterion \eqref{eq:std-objective} is
$\bw^\ast \propto (\bm{\Sigma}^x)^{-1}\bm{\delta}_x \propto
(\bm{\Sigma}^x_{\mathrm{w}})^{-1}\bm{\delta}_x$, i.e. the Fisher
discriminant direction of the regime split, together with
$\bv^\ast \propto (\bm{\Sigma}^y)^{-1}\bm{\delta}_y$, where
$\bm{\Sigma}^x$ and $\bm{\Sigma}^y$ denote the total (mixture)
covariances. The attained value is
\[
\pi(1-\pi)\,|2\pi-1|\,
\bigl(\bm{\delta}_x^\top(\bm{\Sigma}^x)^{-1}\bm{\delta}_x\bigr)
\bigl(\bm{\delta}_y^\top(\bm{\Sigma}^y)^{-1}\bm{\delta}_y\bigr)^{1/2}.
\]
\end{corollary}

The two variants thus target two different, equally interpretable directions: the raw criterion recovers the shift $\bm{\delta}_x$ itself, the standardized criterion its signal-to-noise-optimal counterpart $(\bm{\Sigma}^x_{\mathrm{w}})^{-1}\bm{\delta}_x$, which maximizes the separation of the projected regimes relative to their within-regime spread. When regime \emph{discovery} is the goal, the Fisher direction is the better cut axis whenever the within-regime correlation is strong; the supplementary material quantifies the resulting gains in regime recovery. That superiority is, however, tied to the assumptions of Corollary~\ref{cor:coskew-std}: it requires a within-regime covariance that is (approximately) common across regimes, and enough rows to estimate the whitening transform stably. When each regime carries its own covariance structure, the cancelled terms of Lemma~\ref{lem:coskew-decomp} re-enter and the pooled whitening misweights directions; and in small nodes the inverted covariance injects estimation noise into every score. In those settings the raw criterion is the more robust choice, which is why it remains the default throughout; Section~\ref{sec:examples} illustrates both sides of this boundary empirically.

\begin{proposition}[consistency]\label{prop:coskew-consist}
Let $(\hat{\bw},\hat{\bv})$ maximize the empirical criterion $n^{-1}\sum_{i} t_i^2 u_i$ over unit-norm pairs, computed from $n$ i.i.d.\ draws of the mixture model of
Proposition~\ref{prop:coskew-ident} with $\pi\neq\tfrac12$, nonzero shifts and finite sixth moments. Then, as $n\to\infty$ with $p$ and $q$ fixed, $\hat{\bw}\to\bw^\ast$ (up to sign) and $\hat{\bv}\to\bv^\ast$ almost surely. The same holds for the standardized criterion \eqref{eq:std-objective}, with the targets of Corollary~\ref{cor:coskew-std}.
\end{proposition}

\begin{remark}[sample-size requirements]\label{rem:samplesize}
The empirical criterion fluctuates around \eqref{eq:coskew-ident} at the parametric rate $O_P(n^{-1/2})$, but with constants governed by \emph{sixth}-order moments (the variance of a third-moment average), which is why third-order criteria demand visibly larger samples than second-order ones. Reliable recovery of $\bw^\ast$ requires thepopulation signal
$\pi(1-\pi)\,|2\pi-1|\,\|\bm{\delta}_x\|^2\,\|\bm{\delta}_y\|$ to dominate those fluctuations, so the required $n$ grows like
$(2\pi-1)^{-2}$ as the mixture approaches balance, and no sample size compensates at $\pi=\tfrac12$ exactly. The simulation study in the supplementary material traces this behaviour empirically across the
$(\pi, n)$ grid, and Section~\ref{sec:conclusions} returns to its
practical implications.
\end{remark}

\begin{remark}[behaviour in $\gamma$]\label{rem:gamma}
Under the mixture model the combined objective \eqref{eq:combined-objective} is the weighted sum of two surfaces with distinct peaks,
\[
(1-\gamma)\,\bw^\top\bigl(\bm{\Sigma}^{xy}_{\mathrm{w}}
+\pi(1-\pi)\,\bm{\delta}_x\bm{\delta}_y^\top\bigr)\bv
\;+\;
\gamma\,\pi(1-\pi)(2\pi-1)\,
(\bw^\top\bm{\delta}_x)^2\,(\bv^\top\bm{\delta}_y),
\]
peaking at the leading covariance pair and at the shift pair of Proposition~\ref{prop:coskew-ident} respectively. As $\gamma$ grows, the global maximizer does not rotate continuously from one to the
other: it stays in one basin until the weighted peak heights cross, at a single crossing weight $\gamma^\ast$ determined by the relative magnitudes of the two signals, and then jumps. This is the transition
observed empirically in Section~\ref{sec:coskew-sim}. Sensitivity to $\gamma$ is therefore of \emph{selection} type: within each basin the recovered direction, and any tree grown from it, are nearly constantin $\gamma$, so downstream performance is plateau-shaped and a coarse cross-validation grid resolves the choice completely. Because the two raw terms are moments of different order, $\gamma^\ast$ is not comparable across problems of different score scale; the standardized objective \eqref{eq:std-objective} removes this dependence, as both of its terms are dimensionless.
\end{remark}
}

\subsection{A matrix-free higher-order power method}\label{sec:coskew-hopm}

Each weight extraction in \eqref{eq:coskew-objective} is a best
rank-one symmetric tensor problem posed on the current deflated block;
this section describes the solver for one such extraction and, without
loss of generality, writes $\bX$ for the block it acts on. The
higher-order power method (HOPM) of \citet{DeLathauwer2000hopm} is the
standard tool. A na\"ive
implementation would form the third-moment tensor
$\sum_i \bxi\otimes\bxi\otimes\byi \in \R^{p\times p\times q}$, which is
$O(p^2 q)$ in storage and prohibitive for wide predictor blocks. Instead, a
\emph{matrix-free}, block-symmetric alternating scheme that never
materializes the tensor itself is used.
Fix the current $\bv$ (hence $\bu = \bY\bv$). The objective in
\eqref{eq:combined-objective} is linear in $\bv$, so the maximizing
response weight is the normalized gradient
\begin{equation}
\bv \;\propto\; \bY^\top\!\big[(1-\gamma)\,t + \gamma\,(t\odot t)\big]/n,
\label{eq:v-update}
\end{equation}
where $\odot$ denotes the Hadamard (or element-wise) matrix product. Fixing $\bv$, the
pure-coskewness update for $\bw$ is the leading eigenvector of the
$p\times p$ matrix
\begin{equation*}
\bm{A}(u) = \bX^\top \operatorname{diag}(u)\,\bX/n,
\end{equation*}
obtained by power iteration through the matrix-free product
\begin{equation*}
\bm{A}(u)\bm{z} = \bX^\top\!\big[u \odot (\bX\bm{z})\big]/n;
\end{equation*}
each
step costs $O(n(p+q))$ and no $p\times p$ matrix is ever stored. The
matrix $\bm{A}(u)$ is symmetric but \emph{indefinite}, so the iteration
converges to the eigenvector that corresponds to the eigenvalue of largest \emph{magnitude}.  \rev{Because}
$\E[t^2 u]$ is odd in $\bv$, a negative dominant eigenvalue is resolved
by flipping the sign of $\bv$. For the combined objective
$0<\gamma<1$ the $\bw$-update is a safeguarded normalized-gradient
fixed point that ascends \eqref{eq:combined-objective} monotonically;
it reduces continuously to the covariance power method as
$\gamma \to 0$. The alternation is initialized deterministically from
the covariance leading singular pair, and run from a small number of deterministic starts
(the covariance warm start plus a few coordinate starts at the most
promising variables). Note that this is owed to the ascent being only
locally convergent.  The full procedure is
summarized in Algorithm~\ref{alg:hopm}; a soft-thresholding step
\citep{Serneels2025sparse} may be applied to the converged $\bw$ to
obtain sparse directions, exactly as in the covariance case.

\rev{The standardized objective \eqref{eq:std-objective} requires no
new solver. Substituting $\bm{a} = (\bm{\Sigma}^x)^{1/2}\bw$ and
$\bm{b} = (\bm{\Sigma}^y)^{1/2}\bv$ turns the constraints $\Var(t)=\Var(u)=1$ into $\|\bm{a}\|=\|\bm{b}\|=1$, so the standardized problem is exactly the raw problem \eqref{eq:coskew-objective} posed on \emph{whitened} blocks $\bX(\bm{\Sigma}^x)^{-1/2}$ and $\bY(\bm{\Sigma}^y)^{-1/2}$. The
implementation whitens through the economy singular value decomposition of each block.  This has computational cost $O\bigl(np\min(n,p)\bigr)$ once per extraction, no $p\times p$ matrix is formed.  The decomposition steps have singular values clipped from below at $10^{-8}$ times the largest, so rank-deficient or near-collinear blocks whiten stably; Algorithm~\ref{alg:hopm} then runs unchanged on the whitened pair and the converged direction is mapped back and renormalized. The clipping is the price of the
variant on ill-conditioned blocks: like any CCA-type criterion it inverts a covariance, which is why the raw criterion remains the default for wide or nearly collinear predictor blocks.}

\begin{algorithm}[htb]
\caption{One coskewness direction (matrix-free HOPM), $\bX$-side loop.}
\label{alg:hopm}
\begin{algorithmic}[1]
\Require Centred, scaled $\bX,\bY$; weight $\gamma\in[0,1]$; tolerances.
\State Initialize $(\bw,\bv)$ from the leading singular pair of
       $\bX^\top\bY/n$.
\Repeat
  \State $\bt \gets \bX\bw$;\quad update
         $\bv \propto \bY^\top[(1-\gamma)\bt+\gamma(\bt\odot \bt)]/n$,
         renormalize;\quad $\bu \gets \bY\bv$.
  \If{$\gamma = 1$}
     \State $\bw \gets$ dominant-magnitude eigenvector of
            $\bm{A}(\bu)=\bX^\top\!\operatorname{diag}(\bu)\bX/n$ by
            matrix-free power iteration;
     \State if its Rayleigh quotient is negative, set
            $\bv\gets-\bv,\;\bu\gets-\bu$.
  \Else
     \State $\bw \gets$ safeguarded normalized-gradient fixed point of
            \eqref{eq:combined-objective}.
  \EndIf
\Until{the objective \eqref{eq:combined-objective} stops increasing.}
\State \textbf{return} $\bw$ (optionally soft-thresholded).
\end{algorithmic}
\end{algorithm}

% =====================================================================
\section{The coskewness objective in isolation}\label{sec:coskew-sim}
% =====================================================================

Before embedding the coskewness decomposition in a tree,  this section presents a limited simulation study that attests that the method based on coskewness can recover the inherent structure the covariance objective
cannot, in a controlled setting. A predictor block of six columns is constructed in which a
single column drives the response through its \emph{square}, next to a weak linear decoy. Note that this is \rev{a} purely
third-order relationship. Let $z$ be a
standard-Gaussian variable, $x_2$ a second, independent one, and
$n_1,\dots,n_4$ four independent Gaussian noise columns; set
$\bX = [\,z, x_2, n_1, n_2, n_3, n_4\,]$ and a single response
\begin{equation}
y = (z^2 - 1) + 0.1\,x_2 + 0.05\,\varepsilon .
\label{eq:planted-skew}
\end{equation}
The population moments are decisive: $\operatorname{cov}(z,y)=\E[z^3]=0$
while $\E[z^2 y] = \E[z^4]-\E[z^2] = 2$. The covariance objective is
thus \emph{unable} to rank the planted direction $z$ above the decoy
$x_2$, whereas the coskewness objective sees a strong signal on $z$. Here $z$ is
drawn \emph{antithetically} ($z$ paired with $-z$) so that every odd
sample moment of $z$ vanishes exactly and the finite-sample covariance
$\widehat{\operatorname{cov}}(z, z^2-1)$ is $0$ by construction, leaving
the decoy as the only linear signal; $n = 1000$.

\begin{figure}[htbp]
\centering
\includegraphics[width=0.96\linewidth]{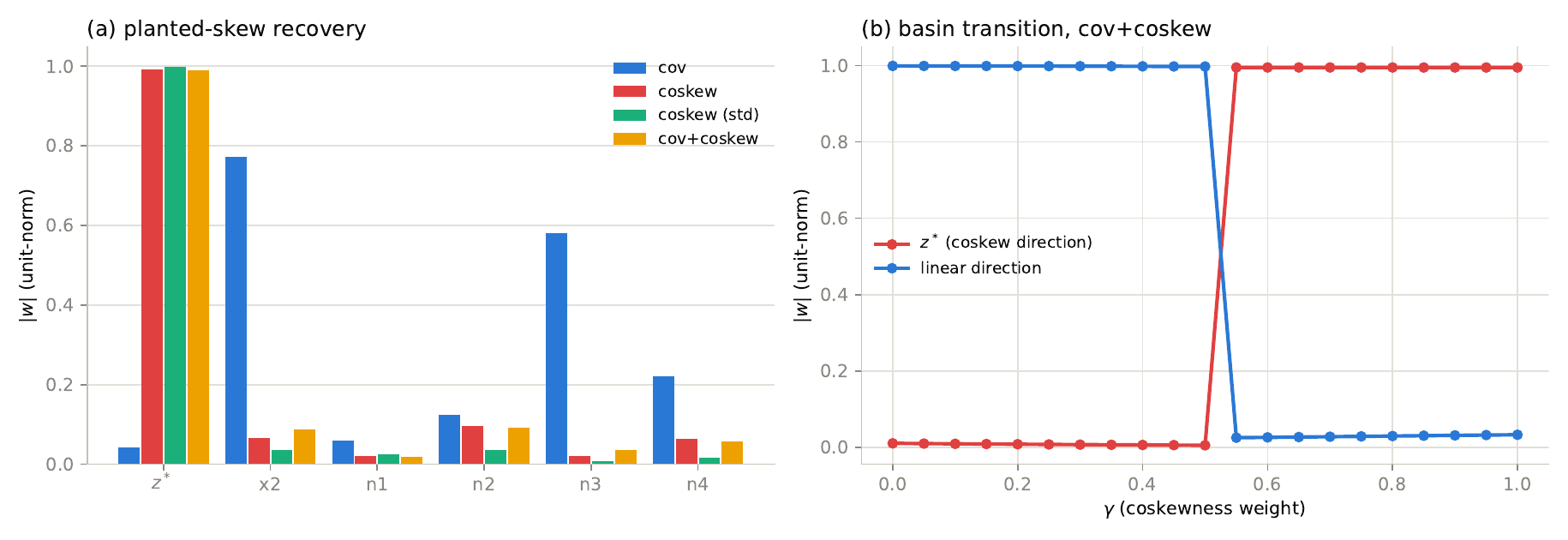}
\caption{The coskewness objective in isolation. (a) First $\bX$-weight
vector (absolute value, unit norm) on the planted-skew design
\eqref{eq:planted-skew}: the covariance objective loads on the linear
decoy $x_2$, while the coskewness and combined objectives recover the
planted direction $z^\ast$\rev{, as does the score-standardized
variant \eqref{eq:std-objective}}. (b) Basin transition of the combined
objective on a companion design with a competing covariance-attractive
direction: as the coskewness weight $\gamma$ grows, the recovered
direction switches sharply from the linear to the skewed axis.}
\label{fig:coskew-planted}
\end{figure}

Fitting a single-component twoblock on the autoscaled data, the
covariance objective places almost no weight on the planted direction
($|w_z| \approx 0.04$) and spreads its energy over the decoy and
whatever noise columns happen to correlate with $y$ in sample, whereas
the coskewness objective loads almost entirely on $z$
($|w_z| \approx 0.99$); Figure~\ref{fig:coskew-planted}(a).
\rev{The standardized variant \eqref{eq:std-objective} behaves
identically here ($|w_z| > 0.99$): the predictor columns of design \eqref{eq:planted-skew} are independent, so the block covariance is near the identity and the Fisher direction of Corollary~\ref{cor:coskew-std} coincides with the raw shift direction. The two variants separate only under within-block
correlation, which the supplementary material examines.} The
antithetic sampling is not cosmetic: with an ordinary Gaussian draw the
sample third moment of $z$ has standard deviation $\sqrt{10/n}\approx0.1$
at $n=1000$, which is the same size as the decoy's coefficient. Therefore, the
covariance objective would ``find'' the skewed direction through pure
sampling noise, thence obscuring the mechanism. Under antithetic sampling the
covariance weight on $z$ stays below $0.06$ across twenty seeds, while
the coskewness weight is consistently near one.

Figure~\ref{fig:coskew-planted}(b) illustrates the combined objective
\eqref{eq:combined-objective} on a companion design in which a
covariance-attractive direction (carrying a mild skewness of its own)
competes with the antithetic coskewness direction. As $\gamma$
increases, the optimum does not blend smoothly between the two: it
\emph{switches basins} sharply, here just above $\gamma \approx 0.5$.
Below the transition the extra coskewness term is absorbed by the
linear direction's own mild skewness; above it, the third-order signal
of the planted direction dominates. This basin structure is why the
local-convergence safeguard and the deterministic multi-start of
Section~\ref{sec:coskew-hopm} matter: a single covariance-warm-started
ascent can remain trapped in the covariance basin even for the pure
coskewness objective, and the coordinate restarts are what free it.
\rev{The sharpness of the transition should not be read as sensitivity of the method to $\gamma$: it concerns the recovered \emph{direction}, which Remark~\ref{rem:gamma} shows to be piecewise
(nearly) constant in $\gamma$, jumping once between two basins. On either side of the jump the direction barely moves, so downstream performance is plateau-shaped in $\gamma$; the supplementary material traces these plateaus on all three examples of this paper, which is what makes the coarse cross-validation grid of Section~\ref{sec:coskew-obj} sufficient in practice.}

% =====================================================================
\section{The twoblock clustering tree}\label{sec:tbtree}
% =====================================================================

The twoblock clustering tree (\tbtree) introduces a novel, deterministic way to build decision trees for multivariate responses that can result in shallow, highly interpretable models. In a nutshell, the twoblock tree accomplishes this by using the first scores of a twoblock model that accounts for coskewness in each node. Results from the previous paragraph, along with well-established results on properties of the third moment, lead us to assume that including a \rev{third-order comoment} into the split impurity criterion is conducive to a more efficient separation of the groups that then become the next-level nodes. In the leaves, though, prediction of the dependent variable is the objective, such that in most cases, twoblock models based on a covariance-alone criterion are preferable.  The default proposed configuration of twoblock clustering trees therefore consists of using twoblock models based on \rev{third-order comoments} as impurity to split the data, but using the regular, \rev{covariance-based} twoblock models as predictive models in the leaves. At this point, it is noted that both split and leaf models can either be dense or sparse\rev{, but} the latter will only be beneficial if truly uninformative variables are present in the data.    Figure~\ref{fig:tbtree-schematic} shows the resulting structure schematically for a single split.

\begin{figure}[htb]
\centering
\begin{tikzpicture}[
    >={Stealth[length=2.6mm]},
    font=\small,
    rootnode/.style={ellipse, draw, thick, align=center,
                     minimum width=52mm, minimum height=20mm, fill=black!4},
    leafnode/.style={rectangle, draw, thick, rounded corners=2pt, fill=black!2},
    edgelab/.style={font=\footnotesize, fill=white, inner sep=1.5pt},
  ]

  % ---- Root node: coskewness twoblock split ----
  \node[rootnode] (root) at (0,0)
    {\textbf{Root node} ($n$ rows)\\[2pt]
     \twoblock\ \emph{coskewness} split\\[2pt]
     score $t_1=\bxi^\top\bw_1$, threshold $\tau$};

  % ---- Split-direction bar chart (coskewness weights w1) ----
  \node[draw, dashed, gray, inner sep=3pt, anchor=west] (splitbox) at (4.0,0) {%
    \begin{tikzpicture}[baseline]
      \coefchart{0/0.72,1/-0.12,2/0.48,3/0.05,4/-0.24,5/0.10}
      \node[font=\scriptsize, anchor=south, gray!50!black] at (1.1,0.82)
        {coskewness split weights $\bw_1$};
    \end{tikzpicture}};
  \draw[gray, dashed] (root.east) -- (splitbox.west);

  % ---- Leaf nodes: covariance twoblock models with coefficient bar charts ----
  \node[leafnode] (leafL) at (-3.6,-4.6) {%
    \begin{tikzpicture}[baseline]
      \node[font=\small, anchor=south, align=center] at (1.1,0.95)
        {\textbf{Leaf} $\ell_1$: covariance \twoblock\\ $\hat\byi=\bxi^\top\bB_{\ell_1}$};
      \coefchart{0/0.66,1/0.22,2/-0.34,3/0.08,4/0.03,5/0.16}
      \node[font=\scriptsize, anchor=north, gray!50!black] at (1.1,-0.35)
        {leaf coefficients $\bB_{\ell_1}$};
    \end{tikzpicture}};

  \node[leafnode] (leafR) at (3.6,-4.6) {%
    \begin{tikzpicture}[baseline]
      \node[font=\small, anchor=south, align=center] at (1.1,0.95)
        {\textbf{Leaf} $\ell_2$: covariance \twoblock\\ $\hat\byi=\bxi^\top\bB_{\ell_2}$};
      \coefchart{0/-0.30,1/0.60,2/0.40,3/-0.52,4/0.18,5/-0.08}
      \node[font=\scriptsize, anchor=north, gray!50!black] at (1.1,-0.35)
        {leaf coefficients $\bB_{\ell_2}$};
    \end{tikzpicture}};

  % ---- Split arrows ----
  \draw[->, thick] (root.south) to
    node[edgelab, pos=0.55]{$t_1<\tau$} (leafL.north);
  \draw[->, thick] (root.south) to
    node[edgelab, pos=0.55]{$t_1\ge\tau$} (leafR.north);

\end{tikzpicture}
\caption{Schematic of a depth-one \tbtree. The root (ellipse) fits a
twoblock model under the \emph{coskewness} split objective; its first
$\bX$-score $t_1=\bxi^\top\bw_1$ orders the rows, and a threshold $\tau$
routes each row to one of two leaves (rectangles). Every leaf carries its
own \emph{covariance} twoblock model $\hat\byi=\bxi^\top\bB_\ell$. The
bar charts show, for a hypothetical six-variable predictor block, the
coskewness split weights $\bw_1$ and the two leaves' regression
coefficients (blue positive, red negative), whose contrast illustrates
the distinct local regimes. This figure is purely illustrative: the
displayed weights and coefficients are specimen values, not fit to any
dataset.}
\label{fig:tbtree-schematic}
\end{figure}

\subsection{Recursive induction}\label{sec:tbtree-induction}

\paragraph{The \rev{decision} tree.}
A tree $\mathcal{T}$ partitions the rows into a set
$\mathcal{L}(\mathcal{T})$ of disjoint leaves $\ell$; on each leaf a
separate twoblock model $(\bW_\ell,\bB_\ell)$ is fit to its
members. The \rev{twoblock clustering} tree grows this tree by recursive top-down induction. At
the root, a twoblock model -- with the objective of choice, covariance
\eqref{eq:cov-objective}, coskewness \eqref{eq:coskew-objective} or the
combination \eqref{eq:combined-objective} -- is fitted on all $n$ rows;
the first score $t_{\cdot 1}$ of that model defines a one-dimensional
ordering of the rows along which a cut $\tau$ is searched. Once a cut is
committed, the rows split into a left child (rows with $t_{i1}<\tau$)
and a right child; on each child a new twoblock model is fit and
the recursion continues until a stopping rule fires.
\rev{Every node's model is fit on that node's own rows only, with both
blocks re-centred and re-autoscaled on those rows, so the moment
objectives of Section~\ref{sec:coskew-obj} are evaluated on variables
standardized relative to the node's own location and spread at every
level of the recursion.}

The decomposition objective enters a node in two distinct roles, and
\tbtree\ allows them to differ. In its \emph{splitting} role it shapes
the first score $t_{\cdot 1}$ along which the cut is searched: with the
covariance objective $t_1$ is the direction of maximal joint
$\bX$--$\bY$ covariance and \tbtree\ reduces essentially to a twoblock
analogue of PLS-Trees \citep{ErikssonTryggWold2009}; with the coskewness
objective $t_1$ is instead the direction of maximal joint skewness,
which is assumed to be
the direction along which two regimes are most cleanly separated. In its
\emph{leaf} role the fitted model $(\bW_\ell,\bB_\ell)$ supplies the
prediction $\hat{\byi}=\bxi^\top\bB_\ell$. These two roles have opposing
ideal objectives: separating regimes is a third-moment property, so the
split benefits from coskewness (or the combined objective), whereas
predicting inside a leaf is an ordinary regression, for which a
skewness-aligned direction is a poor axis -- it deliberately trades away
the within-regime covariance that a leaf regression needs.

The recommended, and throughout
Sections~\ref{sec:tbtree-sim}--\ref{sec:examples} \emph{default},
configuration therefore {\em decouples} the two: the tree is grown and
routed with a coskewness or combined \emph{split} objective, while every
\emph{leaf} model $\bB_\ell$ is fit under the plain covariance objective.
Operationally the tree structure and its routing directions $\bw_1$ are
fixed first, and each leaf is then refit under the covariance objective on
its own members; the internal routing, computable from $\bX$ alone at
prediction time, keeps the skewness-aligned split. This decoupled tree
retains the regime separation of the third-moment split while paying none
of its predictive cost, and it is the form intended whenever a coskewness or cov+coskew
\tbtree\ is referred to in the benchmarks below.
\rev{In the vocabulary of the goal statement of
Section~\ref{sec:intro}: the split objective serves regime alignment,
the leaf objective serves the predictive constraint, and decoupling is
the device that lets \tbtree\ pursue both.  The recovery-versus-prediction
trade-off is thus a property of the \emph{coupled} variant, largely
designed away in the recommended form.} Using the
coskewness objective in the leaves as well is an option, but is preferable only in niche
settings where the leaf directions are themselves the object of study.
For instance, reading a leaf's third-moment loadings to see which
variables drive that regime's asymmetry could be a niche setting in which such a {\em coupled} variant would be preferred.  Howbeit, on every predictive benchmark
investigated here, it was dominated by the decoupled variant.

\paragraph{Stopping rules.}
A node becomes a leaf if (i) its depth reaches a user-set $d_{\max}$,
(ii) it contains fewer than $2n_{\min}$ rows, where $n_{\min}$ is the
minimum enforced on each child, or (iii) the cross-validation gate of
Section~\ref{sec:cv-gate} refuses the candidate split. Otherwise the
cut minimizing the chosen impurity is accepted and the recursion
continues.

\paragraph{Prediction.}
For a new row, \tbtree\ routes it from the root downward by computing
the first $\bX$-score of each node's fitted model and comparing it to
that node's stored cut, until a leaf is reached; the leaf's twoblock
model then gives $\hat{\byi} = \bxi^\top \bB_\ell$. The routing uses
$t_1 = \bxi^\top\bw_1$, which is computable from $\bX$ alone at
prediction time; this is why the cut is searched on the first
$\bX$-score rather than on a response score or a later component. It
also keeps the exhaustive search linear in $n$ and the leaves
interpretable as intervals of a single latent score.

\subsection{Impurity criterion}\label{sec:impurity}

Given the score ordering, a cut is scored by an impurity that combines,
following \citet{ErikssonTryggWold2009}, a within-child response term,
a within-child $t_1$-homogeneity term and a balance penalty, with
weights $A$ and $B$ (defaults $A=B=0.3$). The default response
term adopted herein is Eriksson's within-child response variance. For problems with
many responses compressed into \rev{a few} $\bY$-side components a
\emph{Y-latent variance} is additionally provided that scores cuts in the reduced
$\bU$-subspace; it coincides with the variance criterion when the number
of retained $\bY$-components matches the response rank and is a genuine
reduction otherwise. This criterion, together with a slope-contrast
objective, a kurtosis-based bimodality criterion and an
impurity-invariant $k$-means partition in the joint $(t_1,u_1)$ score
plane, is catalogued in the Supplementary Material.
\rev{To position those supplementary criteria unambiguously: they are
implementation options within the same cut-search and routing
machinery, i.e. an axis orthogonal to this paper's central contribution,
the third-moment split objective. Thusly, they are not competing methodological
proposals. Each targets one specific regime signature that the
variance criteria are structurally unable to reward (slope
differences, response bimodality, and joint-plane separation,
respectively), each carries at most the balance weights $A,B$ as
tuning parameters, and none displaces the variance criteria as the
default. The supplementary material motivates each criterion, states
its tuning defaults, and compares all four criteria across the
benchmarks of this paper.} Two cut-search
heuristics are available: \emph{exhaustive} search scans every interior
break of the $t_1$-order (the default), and a \emph{polynomial} search
exploits the piecewise-quadratic structure of the variance impurity for
a search cost independent of $n$. The $k$-means partition, which
clusters the rows in the $(t_1,u_1)$ plane and snaps the resulting
boundary back to a threshold on $t_1$, is selected by a separate flag;
it turns out to pair particularly well with the coskewness objective for
regime recovery (Section~\ref{sec:tbtree-sim}).

\subsection{Cross-validation termination gate}\label{sec:cv-gate}

The greedy rule above accepts any algorithmically valid split, which on
small or noisy nodes can encourage growth that does not generalize. A
cross-validation gate that refuses such splits is added. For a node with
rows $\mathcal{N}$, the node's CV score under a candidate parameter
tuple is the $K$-fold mean of per-response variance-normalized mean
squared errors, summed across responses,
\begin{equation}
S(\mathcal{N};\bm{\sigma}^2) =
\tfrac{1}{K}\sum_{k=1}^K \sum_{j=1}^{q}
\frac{\frac{1}{|V_k|}\sum_{i\in V_k}(y_{ij}-\hat{y}_{ij})^2}{\sigma_j^2}.
\label{eq:cv-score}
\end{equation}
Given a candidate cut partitioning $\mathcal{N}$ into
$\mathcal{L},\mathcal{R}$, the gate compares the parent score
$S_{\text{p}} = S(\mathcal{N};\bm{\sigma}^2_\mathcal{N})$ with the
size-weighted combined child score
\begin{equation*}
S_{\text{c}} =
\frac{|\mathcal{L}|\,S(\mathcal{L};\bm{\sigma}^2_\mathcal{N})
+|\mathcal{R}|\,S(\mathcal{R};\bm{\sigma}^2_\mathcal{N})}
{|\mathcal{L}|+|\mathcal{R}|}
\end{equation*}
and accepts the split \rev{if and only if}
$(S_{\text{p}}-S_{\text{c}})/S_{\text{p}} \ge \tau_{\text{cv}}$, with a
default $\tau_{\text{cv}}=0.01$. Crucially, both children are scored
with the \emph{parent}'s per-response variance $\bm{\sigma}^2_\mathcal{N}$
as the common denominator: scoring each node against its own variance
makes the scores incomparable across nodes, and on regime-jump data
would spuriously reject the very root split that separates two regimes.

\rev{Four implementation choices complete the specification of the gate: the number of folds $K$, the way the folds are constructed, the extent to which the twoblock model is re-estimated within each fold, and the choice of the threshold $\tau_{\text{cv}}$. Each is discussed in turn. First, regarding the fold count, $K=5$ is used in every experiment of this paper that activates the gate. On small nodes the fold count shrinks to
$\max\{2, \min(K, \lfloor|\mathcal{N}|/2\rfloor)\}$, and when a candidate child is too small to cross-validate at all, or fewer than two folds admit a valid fit, the gate abstains and the split is decided by the impurity and $n_{\min}$ rules alone.  Second, to construct the folds, the folds are plain shuffled $K$-fold partitions of the node's rows, drawn with a fixed seed so that repeated fits are deterministic; no stratification is attempted, as regime labels are precisely what the tree does not know. Third, every quantity is re-estimated within every fold: for each fold (and each candidate parameter tuple, when per-node parameter selection is active) a complete twoblock model is refit from scratch on the fold's training portion and evaluated on its held-out portion.  Note that this entails per-node centring and autoscaling, weight and loading vectors, and regression coefficients. Nothing is warm-started or reused across folds, so \eqref{eq:cv-score} is an honest out-of-sample estimate at the node level; the same score also drives the per-node selection of component counts and sparsity, so gate and model selection share one criterion. Finally, a few considerations on the threshold.  Because parent and children share the parent's variance denominator, $S$ is a scale-free, per-response-normalized error, and $\tau_{\text{cv}}$ is interpretable across datasets: the
default $\tau_{\text{cv}}=0.01$ demands at least a one-percent relative reduction in normalized CV error before the local model count is doubled. This is a parsimony threshold in the spirit of cost-complexity pruning and the one-standard-error rule of \citet{Breiman1984}. Setting $\tau_{\text{cv}}<0$ disables the gate (every CV-scorable split is accepted). The supplementary material reports a sensitivity sweep of $\tau_{\text{cv}}$ on the energy-efficiency benchmark: the fitted tree is \emph{identical} for every threshold from gate-off up to $\tau_{\text{cv}}=0.05$, assuring that each committed split clears the one-percent bar by a wide margin, and only at $\tau_{\text{cv}}=0.1$ does the gate begin to prune, trading a small amount of accuracy for a much smaller tree. The default thus sits well inside a broad stable region, and $\tau_{\text{cv}}$ acts as a coarse safeguard against marginal splits rather than a delicate tuning parameter.}

\subsection{Algorithm and cost}\label{sec:algorithm}

Algorithm~\ref{alg:tbtree} summarizes one recursive step, launched with
$\mathcal{N}=\{1,\dots,n\}$ and $d=0$. Each internal node fits one
twoblock split model (and, under per-node cross-validation, a small
grid of them); with the matrix-free coskewness solver of
Section~\ref{sec:coskew-hopm} the per-node cost stays of the same order
as the covariance fit. Once the structure is fixed, each of the
$|\mathcal{L}(\mathcal{T})|$ leaves is refit once under the covariance
leaf objective -- a cheap ordinary twoblock fit on the leaf's members,
adding a term linear in $n$ overall. A whole tree fits in seconds on the
benchmarks below -- in contrast to the
thirty-minute-to-two-hour budgets reported for mixed-integer optimal
trees \citep{BertsimasDunn2017}.

\begin{algorithm}[htb]
\caption{One recursive step of \tbtree.}
\label{alg:tbtree}
\begin{algorithmic}[1]
\Require Node rows $\mathcal{N}$, depth $d$, hyper-parameters
$(d_{\max}, n_{\min}, A, B, \tau_{\text{cv}})$, base estimator, a
\emph{split} objective (coskew / cov+coskew) and a \emph{leaf} objective
(cov), impurity, search heuristic.
\State Restrict $\bX_\mathcal{N}, \bY_\mathcal{N}$ to $\mathcal{N}$.
\If{$d \geq d_{\max}$ \textbf{or} $|\mathcal{N}| < 2 n_{\min}$}
   \State Fit the leaf model (leaf objective) on
          $(\bX_\mathcal{N},\bY_\mathcal{N})$ and \textbf{return} a leaf.
\EndIf
\State (Optional) select per-node parameters by CV
       \eqref{eq:cv-score}; record $S_{\text{p}}$ with $\mathcal{N}$'s
       variance as denominator.
\State Fit the split model (split objective) on
       $(\bX_\mathcal{N},\bY_\mathcal{N})$; compute $t_{i1}$.
\State Find $\tau^\ast$ minimizing the chosen impurity on $t_{\cdot1}$
       subject to $n_{\min}$; if none exists, fit a leaf model and
       \textbf{return} a leaf.
\State (Optional) CV gate: if
       $(S_{\text{p}}-S_{\text{c}})/S_{\text{p}} < \tau_{\text{cv}}$,
       fit a leaf model and \textbf{return} a leaf.
\State Recurse on the two children at depth $d+1$.
\end{algorithmic}
\end{algorithm}

% =====================================================================
\section{Simulation: recovering piecewise linear regimes}\label{sec:tbtree-sim}
% =====================================================================

\subsection{Design}\label{sec:sim-design}

The following study tests whether the coskewness objective helps
\tbtree\ recover piecewise multivariate linear regimes. A four-regime latent-variable
mixture is constructed. Each regime $k\in\{1,\dots,4\}$ has an
orthonormal pair of $\bX$-loadings $\bP_k\in\R^{p\times2}$ from the QR
decomposition of a Gaussian matrix and a $\bY$-loading pair
$\bQ_k\in\R^{q\times2}$; the regime is centred at a Gaussian location
$\bm{c}_k$. With standard-Gaussian latent scores $\bT_k$, the rows are
\begin{equation}
\bX_k = \bm{c}_k + \bT_k \bP_k^\top + \sigma_\varepsilon \bm{E}_x,
\qquad
\bY_k = \bm{\mu}_k + \bT_k \bQ_k^\top + \sigma_\varepsilon \bm{E}_y,
\end{equation}
with i.i.d.\ Gaussian noise $\bm{E}_x,\bm{E}_y$ at level
$\sigma_\varepsilon$. Setting $\bm{\mu}_k=\bm 0$ gives the slope-only
\emph{Scenario A}; drawing $\bm{\mu}_k$ from a Gaussian of standard
deviation $2$ gives the regime-jump \emph{Scenario B}. Throughout, $n_k=100$, $p=10$, $q=3$ and a $70/30$ split
are used, averaged over \rev{twenty} seeds. Each regime is a distinct local linear map, so the union is a
genuinely piecewise multivariate linear structure -- exactly the target
of the paper's central claim.

Depth-four \tbtree s are grown with two $\bX$-components per node,
comparing the three decomposition objectives (covariance, coskewness,
combined with $\gamma=0.5$) under two cut searches: the exhaustive
variance cut on $t_1$ and the $k$-means partition in the $(t_1,u_1)$
plane. Regime recovery is measured by the adjusted Rand index (ARI) \citep{HubertArabie1985}
between the raw leaf partition and the true regime labels -- the
strictest such measure, as it credits neither extra leaves nor a
majority-vote relabelling -- and predictive quality by the mean test
$R^2$.

\subsection{Results}\label{sec:sim-results}

\begin{figure}[!htb]
\centering
\includegraphics[width=0.85\linewidth]{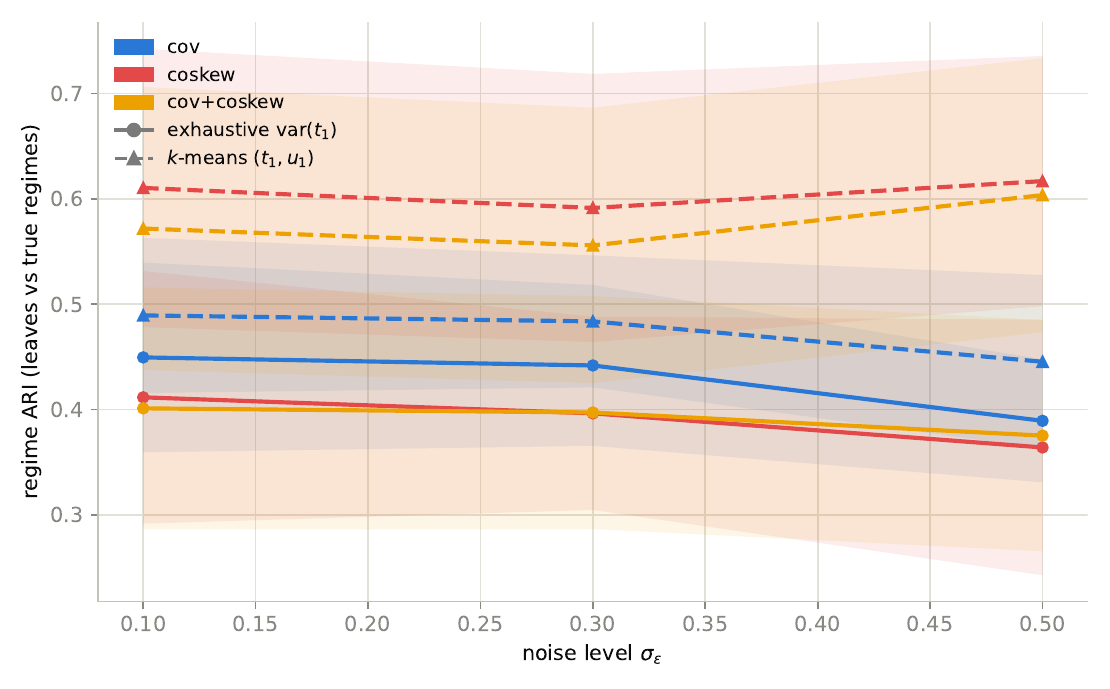}
\caption{Coskewness and regime recovery on the four-regime simulation
(Scenario A): regime-recovery ARI against noise level for the three
objectives (colour) under the exhaustive variance cut (solid) and the
$k$-means $(t_1,u_1)$ split (dashed)\rev{, with shaded bands showing
$\pm 1$ standard deviation over twenty seeds}. The coskewness and combined
objectives with the $k$-means split give the highest ARI, and their
margin over the covariance objective grows with noise.}
\label{fig:coskew-regime}
\end{figure}

Table~\ref{tab:sim-headline} reports the two objectives that bracket the
comparison at each noise level, and Figure~\ref{fig:coskew-regime}
shows the full ARI-versus-noise picture. Three findings stand out.
First, for regime recovery the coskewness objective is clearly superior,
and the effect is strongest with the $k$-means split: in Scenario A the
covariance objective's ARI \emph{falls} from \rev{$0.45$ to $0.39$} as the
noise rises from $\sigma_\varepsilon=0.1$ to $0.5$, whereas the
coskewness $k$-means variant \rev{holds steady at $0.59$--$0.62$} -- the
advantage grows precisely where the covariance signal decays and the
mixture's third-moment signature persists.
\rev{At the same time, an ARI of about $0.6$ is partial alignment,
not recovery: even the best split configuration leaves appreciable
mixing between leaves and regimes, as must be expected when greedy
binary cuts on a single score carve four obliquely oriented regimes
under third-moment sampling noise
(Remark~\ref{rem:samplesize}). The operative claim is therefore the
relative one of the goal statement in Section~\ref{sec:intro}:
substantially better regime alignment than covariance splitting, 
not exact regime identification.} Second, the coskewness trees
are also structurally closer to the truth, using about six leaves for
four regimes against the covariance trees' ten to twelve. Third, this
regime-recovery gain carries a predictive cost on this deliberately
smooth mixture: because each leaf's regression is a linear latent model,
a $t_1$ that maximizes between-regime separation rather than
within-regime covariance sacrifices some $R^2$ (Scenario~A,
$\sigma_\varepsilon=0.3$: \rev{$0.46$} for coskewness against \rev{$0.71$} for
covariance). The combined objective sits between the two.
\rev{This is the recovery-versus-prediction trade-off of the goal
statement in Section~\ref{sec:intro}, and it is worth locating
precisely. It attaches to the \emph{coupled} tree, in which one
skewness-aligned direction serves both routing and leaf regression.
Decoupling recovers much of the gap -- the coskewness-split tree with
covariance leaves reaches $R^2=0.63$ at $\sigma_\varepsilon=0.3$ --
though on this scenario not all of it: Scenario~A is deliberately the
hardest predictive case, a smooth slope-only mixture whose regimes
barely differ in location, so even the routing costs some fit. On the
mean-jump Scenario~B the decoupled tree predicts within two points
of $R^2$ of the covariance tree while recovering regimes better
(ARI $0.49$ against $0.42$ at $\sigma_\varepsilon=0.15$), and on both
real-world benchmarks of Section~\ref{sec:examples} the recommended
decoupled configuration predicts on par with or better than the
covariance tree. Section~\ref{sec:conclusions} returns to this
trade-off.} The same
qualitative ordering holds in the regime-jump Scenario~B, where the
coskewness $k$-means ARI reaches \rev{$0.66$} against the covariance
objective's \rev{$0.42$} at the highest noise level.
\rev{The score-standardized variant of Section~\ref{sec:coskew-obj}
was evaluated on the same grid and degrades both recovery and
prediction here (e.g.\ Scenario~A at $\sigma_\varepsilon=0.5$: $k$-means
ARI $0.33$ against $0.62$ for the raw criterion). As
Section~\ref{sec:coskew-theory} predicts, since this generator gives
every regime its own loading matrix and hence regime-specific
within-regime covariances, precisely the setting in which the pooled
whitening misweights directions and the raw criterion is the robust
choice.}

\begin{table}[htb]
\centering
\caption{Four-regime simulation: regime-recovery ARI (raw leaf
partition vs.\ true regimes) and mean test $R^2$, averaged over
\rev{twenty} seeds, for the covariance objective with the exhaustive variance cut
and the coskewness objective with the $k$-means $(t_1,u_1)$ split -- the
best regime-recovery configuration. The coskewness objective wins on
ARI at every noise level and by a growing margin, at a predictive cost
on this smooth mixture. The combined objective (not shown) interpolates;
the full grid is in Figure~\ref{fig:coskew-regime}.}
\label{tab:sim-headline}
\begin{tabular}{llccc}
\toprule
Scenario & $\sigma_\varepsilon$ &
  \multicolumn{1}{c}{cov, exhaustive} &
  \multicolumn{1}{c}{coskew, $k$-means} & \\
 & & ARI\;/\;$R^2$ & ARI\;/\;$R^2$ & \\
\midrule
A (slope only) & $0.1$ & \rev{$0.45\;/\;0.89$} & \rev{$0.61\;/\;0.63$} & \\
A              & $0.3$ & \rev{$0.44\;/\;0.71$} & \rev{$0.59\;/\;0.46$} & \\
A              & $0.5$ & \rev{$0.39\;/\;0.42$} & \rev{$\bm{0.62}\;/\;0.27$} & \\
\midrule
B (regime jump) & $0.05$ & \rev{$0.42\;/\;0.97$} & \rev{$0.64\;/\;0.87$} & \\
B               & $0.15$ & \rev{$0.42\;/\;0.96$} & \rev{$0.63\;/\;0.85$} & \\
B               & $0.30$ & \rev{$0.42\;/\;0.89$} & \rev{$\bm{0.66}\;/\;0.79$} & \\
\bottomrule
\end{tabular}
\end{table}

Summarizing,  the coskewness objective recovers
the regimes markedly better than the covariance objective that
underlies PLS-Trees for piecewise multivariate linear data, with the advantage widening as noise obscures the
first-order structure. When the goal is regime discovery, the
coskewness objective with the $k$-means split is the configuration to
use; when pure predictive accuracy on smooth data is the goal, the
covariance or combined objective is preferable.

% =====================================================================
\section{Real-world benchmarks}\label{sec:examples}
% =====================================================================

\rev{Real data come without regime labels, so the structural half of
the goal statement of Section~\ref{sec:intro} of regime alignment
cannot be scored directly in this section. The two benchmarks
therefore divide the
burden of proof as follows. Predictive accuracy against strong
black-box references is the \emph{falsifiable} half of the claim: it
verifies the constraint that regime-aligned trees do not purchase
their structure at the price of fit. The structural payoff, which the
simulation of Section~\ref{sec:tbtree-sim} scored directly against
ground truth, surfaces here as the interpretability of the fitted
structure in the form of a dendrogram of operating regimes
(Figure~\ref{fig:energy-dendro}), each an inspectable local linear
model, and on the gas-turbine data a partition of the operating
envelope that tracks physically meaningful conditions. This is why
the tables below emphasize prediction: it is the half of the goal
that real data can adjudicate. The full benchmark protocol is documented in the supplementary
material, including software
versions, the configuration and tuning policy of every competing model, wall-clock fit times, and uncertainty quantification through
twenty repeated splits (energy) and a three-fold rolling-origin evaluation for the gas turbine data set. }

\subsection{UCI energy-efficiency benchmark}\label{sec:energy}

The energy-efficiency dataset of \citet{TsanasXifara2012} comprises
$768$ simulated residential buildings parametrized by eight building
attributes and predicts two continuous energy responses (heating and
cooling load). Four of the eight inputs are discrete, which is the
source of the dataset's regime structure, but the response surface is
otherwise smooth and non-linear rather than sharply piecewise -- the
opposite end of the spectrum from the piecewise mixtures of the
simulation, and the setting in which the paper's second claim is tested,
that \tbtree\ can also deliver state-of-the-art predictive accuracy.

\begin{figure}[htbp]
\centering
\includegraphics[width=0.96\linewidth]{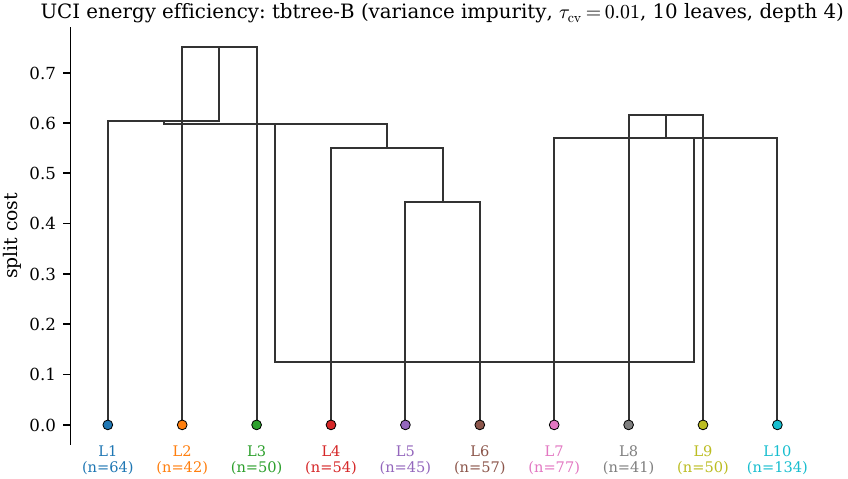}
\caption{UCI energy efficiency: a depth-four \tbtree. The dendrogram
exposes the hierarchical regime structure of the simulated buildings,
with the deepest low-cost split separating the two overall-height
levels. Bracket heights are the impurity costs at which each split was
committed.}
\label{fig:energy-dendro}
\end{figure}

Table~\ref{tab:energy} reports mean test $R^2$ on a fixed $80/20$ split
(seed $0$) with approximate parameter counts, for the two decoupled
\tbtree\ configurations (a coskewness split and a combined split, each
with covariance leaves) against the usual black-box references. The
\emph{combined} split with covariance leaves is the best \tbtree\
configuration obtained, at mean $R^2=0.975$ -- above a deep CART
regression tree ($0.970$, at $613$ leaves), and within noise of a
$200$-tree random forest ($0.980$) and a two-hidden-layer multilayer
perceptron ($0.978$). It reaches this as a single tree of eleven leaves
with a few hundred stored coefficients: one to three orders of magnitude
fewer parameters than the black-box regressors, and fitted in seconds.
\rev{To quantify the split-to-split uncertainty behind these
single-split numbers, the supplementary material repeats the whole
comparison over twenty random $80/20$ splits: the combined-split tree
averages $R^2 = 0.971 \pm 0.014$, a statistical tie with the deep
CART tree ($0.971 \pm 0.005$) at roughly two per cent of its leaf
count, within one standard deviation of the random forest
($0.981 \pm 0.003$), and ahead of the multilayer perceptron, which
proves the least stable competitor ($0.952 \pm 0.024$). The
single-split values of Table~\ref{tab:energy} all lie within one
standard deviation of their repeated-split means.}
The pure-coskewness split with covariance leaves is more parsimonious
($R^2=0.937$, six leaves) but, on data this smooth, does not need the
extra regime sensitivity.
\rev{The approximate parameter counts of Table~\ref{tab:energy} are
obtained by one common rule: the number of real coefficients the
fitted predictor stores. For the flat \twoblock\ and for every
\tbtree\ leaf this is the $p \times q$ coefficient matrix plus the
$q$-vector intercept, i.e.\ $q(p+1) = 18$ numbers per linear model
here, so the flat model counts $18$ and the trees approximately
$18$ times their leaf count ($6$ leaves $\approx 110$, $11$ leaves
$\approx 200$; the routing directions add a further $p+1$ numbers
per internal node, a minor contribution left out of the rounded
counts). CART and the random forest store one $q$-vector response
mean per leaf, giving $2 \times 613 = 1{,}226$ for the CART tree and
the sum over all $200$ forest trees for the random forest. The MLP
count is the total number of weights and biases,
$(8 \cdot 64 + 64) + (64 \cdot 32 + 32) + (32 \cdot 2 + 2) =
2{,}722$. The same rule underlies the model-size statements of
Section~\ref{sec:gasturbine}.} On this benchmark decoupling barely moves the
predictions relative to a fully coupled tree: the leaves are small and
cross-validated to two or three components, so the covariance and
coskewness leaf regressions almost coincide, and the small predictive
edge of the combined objective lives in the \emph{split} rather than the
leaf. The combined split is thus the sweet spot on smooth non-linear
data, just as a pure-coskewness split is on sharply piecewise data.
Figure~\ref{fig:energy-dendro} shows the resulting tree, whose deepest
low-cost split separates the two overall-height levels that dominate the
predictor block.

\begin{table}[htb]
\centering
\caption{UCI energy-efficiency benchmark (Tsanas \& Xifara, 2012): mean
test $R^2$ on the $80/20$ hold-out and approximate parameter count, for
the two decoupled \tbtree\ configurations (coskewness / combined split,
covariance leaves) and the black-box references. The combined-split
\tbtree\ is the best single-tree model and matches the black-box
regressors to within noise at a fraction of their size. Best single tree
in bold.}
\label{tab:energy}
\begin{tabular}{lcc}
\toprule
Model & mean $R^2$ & $\#$params \\
\midrule
flat \twoblock                                    & $0.876$ & $18$ \\
\tbtree\ coskew split, cov leaves                 & $0.937$ & $\sim\!110$ \\
\tbtree\ cov+coskew split, cov leaves             & $\bm{0.975}$ & $\sim\!200$ \\
CART regression tree                              & $0.970$ & $1\,226$ \\
Random forest, $200$ trees                        & $0.980$ & $155\,394$ \\
MLP $(64,32)$, ReLU                               & $0.978$ & $2\,722$ \\
\bottomrule
\end{tabular}
\end{table}

This example illustrates that, for piecewise multivariate linear data, a combined-split \tbtree\ with covariance
leaves allows a shallow, interpretable model to achieve predictions almost as accurate as state-of-the-art
black-box regressors, while only using a tiny fraction of the latter's
parameters.
\rev{Standardizing the split objective does not pay off on this
benchmark: the combined-split tree drops from $R^2=0.975$ to $0.957$,
marginally below the covariance tree, in line with the guidance of
Section~\ref{sec:coskew-theory} -- with $614$ training rows split
across a depth-four tree, the per-node whitening transform is
estimated on subsets too small for its theoretical advantage to
materialize.}

\subsection{Gas-turbine emissions: prediction under regime
drift}\label{sec:gasturbine}

The second benchmark stresses the decoupled tree on a large, genuinely
non-linear regression with a temporal train/test split. The gas-turbine
dataset of \citet{Kaya2019gasturbine} collects $36{,}733$ hourly readings
from a combined-cycle power plant over 2011--2015: nine ambient and
process predictors and two emission responses, carbon monoxide (CO) and
nitrogen oxides (NOx). Training uses 2011--2014 and testing 2015, so the
test year is an out-of-distribution extrapolation under drifting
operating conditions -- the setting in which a black-box regressor's
advantage on i.i.d.\ data is least assured. Because the two responses
differ in scale by roughly a factor of five, the root mean squared
error of prediction (RMSEP) is reported per response together with a
scale-weighted mean,
$\overline{\mathrm{RMSEP}}_w =
\tfrac{1}{2}\sum_{j}\mathrm{RMSEP}_j/\sigma_j$, where $\sigma_j$ is the
test standard deviation of response $j$; this puts the two responses on
an equal footing and tracks a mean coefficient of determination.

Table~\ref{tab:gasturbine} reports the \revdel{two} decoupled \tbtree\
configurations against the black-box references. On the scale-weighted
mean the decoupled \revdel{coskewness-split tree is the best model
overall} \rev{coskewness-split trees
($\overline{\mathrm{RMSEP}}_w = 0.830$; $0.829$ for the
score-standardized variant with $29{,}349$ training rows and
$n_{\min}=500$, this benchmark sits squarely inside the regime where
Section~\ref{sec:coskew-theory} predicts whitening to be safe) lead
every non-sequential reference: they are ahead of} a $200$-tree random
forest ($0.876$), a multilayer perceptron ($0.882$)\rev{,} and Cubist
($0.863$); the combined-split tree is close behind ($0.834$)\rev{,
and only the temporal convolutional network
\citep[TCN;][]{BaiKolterKoltun2018}, a sequence model that consumes a
$24$-hour context window rather than single rows, edges ahead at
$0.813$}. The advantage is sharpest on CO, where
both decoupled trees have the lowest error of any method; on the harder,
drift-dominated NOx response they are competitive with, though not
ahead of, the deepest black-box models. A fully coupled coskewness tree
is markedly worse on NOx, its skewness-aligned leaves generalizing poorly
across the year boundary. This is exactly the gap the covariance leaves close.
\rev{Two robustness checks in the supplementary material qualify and
support these single-split rankings. First, a rolling-origin
evaluation over three temporal folds (test years $2013$--$2015$, each
trained on all preceding years) shows that fold-to-fold variability
($\pm 0.13$--$0.18$ in $\overline{\mathrm{RMSEP}}_w$) dwarfs the
between-model gaps within the leading group the decoupled trees,
the TCN, the random forest and the MLP are statistically
indistinguishable across folds, with the standardized coskewness tree
the most stable of the group ($0.80 \pm 0.13$). Second, the TCN entry
of Table~\ref{tab:gasturbine} was refreshed by refitting under the
documented software stack; neural sequence models are not
bit-reproducible across library versions, which is itself part of the
uncertainty picture the supplementary material quantifies.}

\begin{table}[htb]
\centering
\caption{Gas-turbine emissions (Kaya et al., 2019), temporal split
(train 2011--2014, test 2015): per-response RMSEP and the scale-weighted
mean $\overline{\mathrm{RMSEP}}_w$ (lower is better), with the number of
local leaf models. The \revdel{two} decoupled \tbtree\ configurations
\rev{(including the score-standardized coskewness split)} are shown
against the black-box references. Best in each column in bold.}
\label{tab:gasturbine}
\setlength{\tabcolsep}{4pt}
\begin{tabular}{lcccc}
\toprule
Model & \# regimes & RMSEP CO & RMSEP NOx &
$\overline{\mathrm{RMSEP}}_w$ \\
\midrule
\tbtree\ coskew split, cov leaves      & $16$ & $1.61$ & $10.44$ & $0.830$ \\
\rev{\tbtree\ coskew split (std.), cov leaves} & \rev{$16$} & \rev{$\bm{1.60}$} & \rev{$10.46$} & \rev{$0.829$} \\
\tbtree\ cov+coskew split, cov leaves  & $16$ & $1.62$ & $10.50$ & $0.834$ \\
TCN ($4$ blocks)                       & --   & \rev{$1.76$} & \rev{$\bm{9.32}$} & \rev{$\bm{0.813}$} \\
Cubist ($10$ committees)               & --   & $1.70$ & $10.75$ & $0.863$ \\
Random forest, $200$ trees             & --   & $1.67$ & $11.20$ & $0.876$ \\
MLP \revdel{$(64,32)$}\rev{$(64,32,16)$}, ReLU & --   & $1.71$ & $11.13$ & $0.882$ \\
flat \twoblock                         & --   & $1.77$ & $12.68$ & $0.966$ \\
CART regression tree                   & --   & $2.33$ & $12.96$ & $1.104$ \\
\bottomrule
\end{tabular}
\end{table}

The predictive parity above is bought at no cost to transparency, which is
the point of the example. The decoupled coskewness tree partitions the
operating envelope into $16$ regimes, each carrying its own covariance
twoblock model -- a linear map from the nine process variables to the
two emissions, with fully readable coefficients (all $16$ leaves
non-degenerate). The regimes are genuinely heterogeneous: across them the
dominant driver of CO takes four distinct values (compressor discharge
pressure in most regimes, but the inlet-filter pressure drop, the turbine
exhaust pressure or the turbine-after temperature in specific operating
regions), and the dominant NOx driver takes two. A plant engineer can
thus read off which process variable governs emissions in each operating
regime, an explanation nor the random forest, the MLP, or the TCN can provide.

Summarizing,  the decoupled \tbtree\ matches or beats every
\revdel{black-box baseline} \rev{row-wise black-box baseline, and is
statistically indistinguishable from the sequence-model TCN across
temporal folds,} on scale-weighted error while remaining a single tree
of $16$ interpretable local emission models; the covariance leaves are
what let the coskewness split compete predictively, whereas a fully coupled
coskewness tree would trail in terms of predictive power.

% =====================================================================
\section{Discussion and outlook}\label{sec:conclusions}
% =====================================================================

This paper has introduced \tbtree, a shallow, deterministic and interpretable, top-down
clustering tree that carries a local (dense or sparse) twoblock dimension reduction
model in every leaf. To the author's knowledge it
is the first tree-based regression method to do so, generalizing the PLS-Trees
of \citet{ErikssonTryggWold2009} from partial least squares to the
joint predictor--response reduction of \citet{Cook2023}. Its central
methodological ingredient is a novel twoblock decomposition whose
latent directions maximize the \emph{coskewness} of the predictor and
response scores rather than their covariance, solved by a matrix-free
higher-order power method that never forms a moment tensor. The
motivation is a simple statistical observation: a node that straddles
two regimes is a mixture, mixtures are non-normal, and the
between-regime separation surfaces in the third-order joint moments
before it dominates the second-order ones.

The empirical study confirms the positioning. A controlled simulation
shows the coskewness objective recovering a planted skewed direction
that the covariance objective is provably blind to. Inside the tree, on
a four-regime piecewise-linear simulation, the coskewness objective
recovers the regimes more faithfully than the covariance objective, with
the advantage widening as noise erodes the first-order signal. On two real-world regressions
the decoupled tree, which consists of a coskewness or combined split with covariance
leaves, then delivers competitive prediction. On the energy-efficiency
benchmark the combined-split tree gives the best single-tree accuracy
observed ($R^2=0.975$), matching a random forest and a multilayer
perceptron to within noise at one to three orders of magnitude fewer
parameters\rev{; across twenty repeated splits it ties a deep CART
regression tree ($0.971$ for both) at two per cent of the latter's
leaf count}. On the gas-turbine emissions benchmark, under a temporal
split with genuine operating-condition drift, the coskewness-split tree
\revdel{attains the lowest scale-weighted error of any method -- ahead
of a random forest, a multilayer perceptron and a temporal
convolutional network --} \rev{leads every row-wise reference on
scale-weighted error. It is ahead of a random forest, a multilayer
perceptron and Cubist, and statistically indistinguishable across
temporal folds from a temporal convolutional network that consumes a
$24$-hour context window,} while remaining a single tree of sixteen
interpretable local emission models.
\rev{This division of labour also answers what \tbtree\ primarily
\emph{is}: a method for regime-aligned, interpretable
piecewise-linear modelling, constrained to remain predictively
competitive.  It is not a predictor {\em per se}, and not an unweighted
compromise between the two goals. Its structural claim is the
relative one established in Section~\ref{sec:tbtree-sim}: partitions
substantially more faithful to the latent regimes than
covariance-based splitting yields, not exact regime recovery.}
Decoupling is what reconciles the two goals: the
third-moment split supplies regime-aware routing while the covariance
leaves supply predictive sharpness, so on the predictive benchmarks the
coskewness split costs little or nothing relative to a covariance tree
yet retains its regime structure. The covariance and coskewness roles
thus occupy complementary ends of a single interpretable dial: a
coskewness (or combined) split for regime-aligned structure, covariance
leaves for predictive accuracy.

In spite of the good results reported, it is noted that \rev{third-order moments} are estimated with more sampling variability than second
moments, so the coskewness objective needs a reasonable sample size to
pay off. On a small concrete-slump benchmark that merely consists of $78$ training rows \citep{YehSlump2007} it
degraded rather than improved both recovery and prediction; the covariance objective remains the right default on small-$n$ problems.
\rev{Remark~\ref{rem:samplesize} gives this limitation its formal
shape -- the sample size required for the third-moment signal to dominate its sixth-order-moment sampling noise grows like
$(2\pi-1)^{-2}$ as the regime mixture approaches balance -- and the
simulation study in the Supplementary Material charts, across the
$(\pi,n)$ grid, where the coskewness objectives overtake the covariance-based state of the art and where they do not.}
The regime-recovery gains on piecewise data also come with a predictive
cost that only the combined objective and the $k$-means split partly
mitigate; choosing $\gamma$ and the split by cross-validation \rev{over
the coarse grid of Section~\ref{sec:coskew-obj}} for a
given dataset is the pragmatic recommendation.  \rev{Note that this is sufficient by the
plateau structure of Remark~\ref{rem:gamma}.}
\rev{The same holds for the choice between the raw and the
score-standardized criterion: the standardized variant is the stronger
regime finder exactly under the conditions of Corollary~\ref{cor:coskew-std}: an approximately common within-regime covariance, strong within-block correlation, and nodes large enough to whiten stably, as on the gas-turbine benchmark and in the supplementary
study. Howbeit, the raw criterion is the robust default whenever regimes carry their own covariance structure or nodes are small, as in the
simulation of Section~\ref{sec:tbtree-sim} and on the
energy-efficiency benchmark.}

Several extensions are natural. The coskewness objective is presently
available only with the dense and sparse twoblock bases; extending it
to the robust and cellwise-robust bases
\citep{Serneels2025rtb,Serneels2026crtb} would combine regime discovery
with outlier resistance. A plane search in
$c\,t_2+(1-|c|)\,t_1$ along the lines of
\citet{ErikssonTryggWold2009}'s \S5.4.1 would relax the axis-aligned
cut. Finally, higher-order objectives beyond coskewness -- cokurtosis,
or a joint independent-component criterion \citep{DeLathauwerICA2000} --
are a natural continuation of the same programme.
\rev{Proposition~\ref{prop:coskew-ident} sharpens the case for the
cokurtosis member of that family: balanced ($\pi=\tfrac12$) mixtures
are invisible to every third-moment criterion, while their bimodality
survives in the fourth-order comomoments.}

% =====================================================================
\section{Software availability}\label{sec:software}
% =====================================================================

\tbtree\ and the coskewness twoblock decomposition are implemented in
the open-source Python package \texttt{twoblock}, with source on GitHub
at \url{https://github.com/SvenSerneels/twoblock}; both have been
publicly available since release 1.0.0 of the package. The standardized
coskewness variant of Section~\ref{sec:coskew-theory} (the
\texttt{standardize} option) and the scripts behind the supplementary
studies are included in the current development version. Reproducible
Jupyter notebooks for the coskewness simulation, the four-regime tree
simulation and both real-world examples ship under \texttt{examples/}.

% =====================================================================
% arXiv build: no Springer "Statements and Declarations" block; code and
% data availability are covered by the Software availability section.
% =====================================================================
\rev{\appendix
\section{Proofs}\label{app:proofs}

\paragraph{Proof of Lemma~\ref{lem:coskew-decomp}.}
Condition on the regime label. Within regime $k$ write
$\bm{x} = \bm{\mu}^x_k + \bm{\varepsilon}$ and
$\bm{y} = \bm{\mu}^y_k + \bm{\eta}$, where $(\bm{\varepsilon},
\bm{\eta})$ has mean zero, covariances $(\bm{\Sigma}^x_k,
\bm{\Sigma}^y_k, \bm{\Sigma}^{xy}_k)$ and vanishing third central
moments. With $e = \bw^\top\bm{\varepsilon}$ and
$h = \bv^\top\bm{\eta}$, so that $t = a_k + e$ and $u = b_k + h$,
\begin{align*}
\E[\,t^2u \mid k\,]
&= \E\bigl[(a_k^2 + 2a_k e + e^2)(b_k + h)\bigr] \\
&= a_k^2 b_k \;+\; 2a_k\,\E[eh] \;+\; b_k\,\E[e^2] \;+\; \E[e^2h] \\
&= a_k^2 b_k \;+\; 2a_k\,\bw^\top\bm{\Sigma}^{xy}_k\bv
   \;+\; b_k\,\bw^\top\bm{\Sigma}^x_k\bw ,
\end{align*}
using that all terms linear in $(e,h)$ have mean zero and that
$\E[e^2h]$ is a third central moment, hence zero. Averaging over $k$
with weights $\pi_k$ gives \eqref{eq:coskew-decomp}. \hfill$\square$

\paragraph{Proof of Proposition~\ref{prop:coskew-ident}.}
With common within-regime covariances the first sum of
\eqref{eq:coskew-decomp} factorizes as
$(\bw^\top\bm{\Sigma}^x_{\mathrm{w}}\bw)\sum_k\pi_k b_k
+ 2(\bw^\top\bm{\Sigma}^{xy}_{\mathrm{w}}\bv)\sum_k\pi_k a_k = 0$,
since the blocks are centred at the mixture mean, i.e.\
$\sum_k \pi_k a_k = \sum_k \pi_k b_k = 0$. Centring also gives
$\bm{\mu}^x_1 = -(1-\pi)\bm{\delta}_x$ and
$\bm{\mu}^x_2 = \pi\bm{\delta}_x$ (and likewise for $\bm{y}$), so
with $d_x = \bw^\top\bm{\delta}_x$ and $d_y = \bv^\top\bm{\delta}_y$,
\[
\sum_k \pi_k a_k^2 b_k
= \pi\,(1-\pi)^2 d_x^2\,\bigl(-(1-\pi)d_y\bigr)
+ (1-\pi)\,\pi^2 d_x^2\,\pi d_y
= \pi(1-\pi)(2\pi-1)\,d_x^2\,d_y ,
\]
which is \eqref{eq:coskew-ident}. The criterion factorizes over
$\bw$ and $\bv$: by Cauchy--Bouniakovsky--Schwarz, $d_x^2$ is maximal over the unit
sphere exactly at $\bw = \pm\bm{\delta}_x/\|\bm{\delta}_x\|$, and
$(2\pi-1)d_y$ at
$\bv = \operatorname{sign}(2\pi-1)\,\bm{\delta}_y/\|\bm{\delta}_y\|$,
with the stated maximal value. Since
$\pi(1-\pi)\neq 0$, expression \eqref{eq:coskew-ident} is the zero
function of $(\bw,\bv)$ if and only if $2\pi-1=0$ or
$\bm{\delta}_x=\bm{0}$ or $\bm{\delta}_y=\bm{0}$. \hfill$\square$

\paragraph{Proof of Corollary~\ref{cor:coskew-std}.}
Dividing \eqref{eq:coskew-ident} by
$\Var(t)\sqrt{\Var(u)} =
(\bw^\top\bm{\Sigma}^x\bw)\,(\bv^\top\bm{\Sigma}^y\bv)^{1/2}$, the
standardized criterion separates into
\[
\pi(1-\pi)(2\pi-1)\;
\frac{(\bw^\top\bm{\delta}_x)^2}{\bw^\top\bm{\Sigma}^x\bw}\;
\frac{\bv^\top\bm{\delta}_y}{(\bv^\top\bm{\Sigma}^y\bv)^{1/2}} .
\]
The $\bw$-factor is a generalized Rayleigh quotient, maximal at
$\bw \propto (\bm{\Sigma}^x)^{-1}\bm{\delta}_x$ with value
$\bm{\delta}_x^\top(\bm{\Sigma}^x)^{-1}\bm{\delta}_x$; the
$\bv$-factor is maximal (with the orientation fixed by the sign of
$2\pi-1$) at $\bv \propto (\bm{\Sigma}^y)^{-1}\bm{\delta}_y$ with
value $(\bm{\delta}_y^\top(\bm{\Sigma}^y)^{-1}\bm{\delta}_y)^{1/2}$.
Finally $\bm{\Sigma}^x = \bm{\Sigma}^x_{\mathrm{w}} +
\pi(1-\pi)\bm{\delta}_x\bm{\delta}_x^\top$, and the
Sherman--Morrison identity gives
$(\bm{\Sigma}^x)^{-1}\bm{\delta}_x \propto
(\bm{\Sigma}^x_{\mathrm{w}})^{-1}\bm{\delta}_x$. \hfill$\square$

\paragraph{Proof of Proposition~\ref{prop:coskew-consist}.}
Let $M_n(\bw,\bv) = n^{-1}\sum_i (\bw^\top\bxi)^2(\bv^\top\byi)$ and
let $M(\bw,\bv)$ denote its population counterpart
\eqref{eq:coskew-ident}. The parameter set, the product of the two
unit spheres, is compact; $(\bw,\bv)\mapsto
(\bw^\top\bm{x})^2(\bv^\top\bm{y})$ is continuous for every
$(\bm{x},\bm{y})$ and dominated by the integrable envelope
$\|\bm{x}\|^2\|\bm{y}\|$, so the class is Glivenko--Cantelli and
$\sup_{\bw,\bv}|M_n - M| \to 0$ almost surely
\citep[Example~19.8]{vanderVaart1998}. By
Proposition~\ref{prop:coskew-ident}, $M$ is continuous with, modulo
the sign of $\bw$, a unique and well-separated maximizer, so the
argmax theorem \citep[Theorem~5.7]{vanderVaart1998} yields the claim.
For the standardized criterion the same argument applies to the
ratio, whose denominator is bounded away from zero on the spheres by
the smallest eigenvalues of $\bm{\Sigma}^x$ and $\bm{\Sigma}^y$; the
finite sixth moments guarantee in addition the $O_P(n^{-1/2})$
fluctuation rate quoted in Remark~\ref{rem:samplesize}.
\hfill$\square$
}

% =====================================================================
% SUPPLEMENTARY MATERIAL (arXiv build): the body of supplementary_R1.tex,
% appended verbatim and numbered S1, S2, ... with its own table, figure
% and equation counters. Cross-references to the main text resolve
% directly (no xr/externaldocument needed).
% =====================================================================
\clearpage
\setcounter{section}{0}
\setcounter{subsection}{0}
\setcounter{table}{0}
\setcounter{figure}{0}
\setcounter{equation}{0}
\renewcommand{\thesection}{S\arabic{section}}
\renewcommand{\theHsection}{S\arabic{section}}
\renewcommand{\thetable}{S\arabic{table}}
\renewcommand{\theHtable}{S\arabic{table}}
\renewcommand{\thefigure}{S\arabic{figure}}
\renewcommand{\theHfigure}{S\arabic{figure}}
\renewcommand{\theequation}{S\arabic{equation}}
\renewcommand{\theHequation}{S\arabic{equation}}
\phantomsection
\addcontentsline{toc}{section}{Supplementary Material}
\begin{center}
{\LARGE\bfseries Supplementary Material}\\[8pt]
{\large to ``Twoblock clustering trees with coskewness-based dimension
reduction: recovering piecewise multivariate linear regimes''}
\end{center}
\bigskip

% =====================================================================
\section{Other impurity criteria}\label{sec:impurity-supp}
% =====================================================================

The main text uses Eriksson's response variance as \tbtree's default
split criterion, together with the optional Y-latent variance $\Phi^u$
of Section~\ref{sec:impurity} of the main paper. These criteria concern
how a candidate cut on the first score $t_1$ is \emph{scored}, an axis
orthogonal to the choice of \emph{split objective} (covariance versus
coskewness) studied in the main text; unless stated otherwise the score
$t_1$ used here is the covariance \twoblock\ score of \citet{Cook2023}.
Three further criteria, all sharing the same cut-search and
prediction-time routing machinery, are described here together with a
concise empirical comparison on the datasets used in this work.

\rev{To position these criteria unambiguously: they are
\emph{implementation options} shipped with the software, secondary to
the paper's methodological contribution (the third-moment split
objective), and none is proposed as a new default. They are
catalogued here -- with their failure cases as well as their wins --
for three reasons. First, each targets a distinct regime signature
that the default variance criteria are structurally unable to
reward. The slope-contrast impurity detects regimes that differ in
their local \emph{slopes} rather than in their response levels, the
Chow-test logic that Scenario~A of
Section~\ref{sec:scenA} isolates. The kurtosis-based impurity
detects \emph{bimodality} of the reduced response scores; it is the
fourth-moment companion of the main paper's third-moment split
objective, driven by the same observation that a node straddling two
regimes is a non-normal mixture, and it connects to the cokurtosis
extension sketched in the outlook of
Section~\ref{sec:conclusions} of the main paper. The $k$-means
partition addresses separations that live in the joint $(t_1,u_1)$
relation rather than in $t_1$ alone; it is the one option that has
graduated into the main text, where it pairs with the coskewness
objective in Sections~\ref{sec:tbtree-sim}
and~\ref{sec:examples}. Second, their tuning burden is deliberately
minimal: the slope-contrast divergence weight has the principled
default $\gamma = 0$, backed by the Chow-decomposition algebra
below, with only a small ridge $\lambda$ for numerical stability;
the kurtosis criterion reuses the main text's balance weights $A,B$
unchanged; and the $k$-means partition is tuning-free. Third, the
cross-dataset summary of Section~\ref{sec:impurity-supp-summary}
supplies the comparison with the default criterion and the
resulting guidance: $\Phi^u$ remains the multi-purpose default, and
a supplementary criterion is worth switching on only when its
target signature is expected in the data -- slope-only regimes,
compositional or bimodal response clusters, or joint-plane
separations, respectively.}

% ---------------------------------------------------------------------
\subsection{Slope-contrast impurity}\label{sec:slope-supp}
% ---------------------------------------------------------------------

Let $\bT_L, \bT_R$ denote the parent's score matrix restricted to each
child of a candidate cut $\tau$. The slope-contrast impurity is
\begin{equation}
\Phi^{\text{slope}}(\tau) \;=\;
\bigl\|\bY_L - \bT_L\hat{\bm{\beta}}_L\bigr\|_F^2 \;+\;
\bigl\|\bY_R - \bT_R\hat{\bm{\beta}}_R\bigr\|_F^2 \;-\;
\gamma\;n_L n_R\,\bigl\|\hat{\bm{\beta}}_L - \hat{\bm{\beta}}_R\bigr\|_F^2,
\label{eq:slope-impurity-supp}
\end{equation}
where
$\hat{\bm{\beta}}_L = (\bT_L^\top \bT_L + \lambda I)^{-1}\bT_L^\top \bY_L$
is a ridge regression of $\bY$ on the \emph{parent}'s score matrix on
the left rows, and similarly for $R$, and $\gamma \geq 0$ controls a
sample-size-weighted reward for between-child slope divergence. A Chow
algebra argument shows that, when the divergence term is weighted by
$n_Ln_R/(n_L+n_R)$ rather than $n_Ln_R$, \rev{Equation}~\ref{eq:slope-impurity-supp}
reduces algebraically to pure within-child residual minimisation;
$\gamma = 0$ is therefore the principled default, and a strictly
positive $\gamma$ is exposed only as a research knob.

Empirically the slope-contrast impurity favours \emph{leaf-to-regime
correspondence} over predictive $R^2$. On the four-regime simulation
Scenario~B of Section~\ref{sec:sim-design} of the main paper
it lifts regime-$F_1$ to $0.745$ (against $\Phi^u$'s $0.679$ under the
default pruning gate) at the cost of a one-percentage-point gain in
$R^2$. On the energy-efficiency benchmark it under-performs at
$R^2 = 0.708$, because the regime boundaries there are aligned with
categorical-level shifts in the response means -- the signal that the
variance terms in $\Phi^u$ capture directly.

% ---------------------------------------------------------------------
\subsection{Kurtosis-based impurity}\label{sec:kurtosis-supp}
% ---------------------------------------------------------------------

A second variant replaces the variance terms in $\Phi^u$ by their
fourth-order analogues. Let $\bU_L, \bU_R$ denote the $K$-column
Y-side latent block restricted to each child. The kurtosis-based
impurity is
\begin{align}
\Phi^{\text{kurt}}(\tau) =
&\;-\Big[\tfrac{|L|}{n}\,\mathrm{kurt}(\bU_L) +
       \tfrac{|R|}{n}\,\mathrm{kurt}(\bU_R)\Big]\\
&\;+\;A\,\Big[\tfrac{|L|}{n}\,\Var(t_{L,1}) +
              \tfrac{|R|}{n}\,\Var(t_{R,1})\Big]
   \;+\;B\,\Big|\tfrac{|L|}{n} - \tfrac{|R|}{n}\Big|,
\nonumber
\label{eq:kurtosis-impurity}
\end{align}
where $\mathrm{kurt}(\bU_\cdot) = \sum_{a=1}^K
\hat{\mu}_4(\bU_{\cdot,a}) / \hat{\sigma}^2(\bU_{\cdot,a})^2$ is the
sum of per-column raw kurtoses. The leading minus sign flips the
optimisation direction: minimising $\Phi^{\text{kurt}}$ maximises the
within-child kurtosis of $\bU$, which drives cuts toward peaked,
unimodal-per-child distributions. The motivation is that a parent
node containing two compositionally distinct regimes will exhibit a
bimodal $\bU$-distribution; isolating each regime in its own child
produces unimodal distributions of higher kurtosis.

Empirically, the kurtosis criterion delivers the most striking single
gain of all impurity variants on the archaeological-glass benchmark of
\citet{Lemberge2000glass} -- an EPXMA/$\mu$-XRF calibration of
\rev{16th--17th-century} glass vessels, used here as an additional benchmark not
reported in the main text:
test-set PbO RMSEP drops from Eriksson's $0.367$ to $0.262$ (a $29\%$
reduction), and the per-response $R^2$ rises from $0.906$ to $0.952$.
The mechanism is that Pb-bearing samples (the high-PbO glasses)
project to a separate $\bU$-cluster from the lead-free majority, so
the bimodality that $\Phi^{\text{kurt}}$ is built to detect is
exactly what determines the PbO response. The trade-off is on the
classical Na$_2$O / K$_2$O response: at scope~B,
$\Phi^{\text{kurt}}$'s root cut at $t_1 = 25.6$ over-includes
non-sodic types in the left child, lifting the Na$_2$O error
slightly. On the simulation -- which uses Gaussian-with-mean-shift
mixtures whose $\bU$-distributions are not bimodal in either child --
$\Phi^{\text{kurt}}$ under-performs the variance criteria by $6$--$8$
percentage points of $R^2$.

% ---------------------------------------------------------------------
\subsection{$k$-means partition in the joint $(t_1, u_1)$ plane}\label{sec:kmeans-supp}
% ---------------------------------------------------------------------

The three impurity criteria above all score a candidate cut on the
first $\bX$-score $t_1$. A fourth option bypasses the impurity score
altogether and lets the cut be chosen by clustering in the joint score
plane. At a node, the pairs
$\{(t_{i1}, u_{i1})\}_{i\in\mathcal{N}}$ -- the first $\bX$-score and
the first $\bY$-score of the node's \twoblock\ model -- are
standardised and partitioned by $k$-means with $k = 2$. Each row is
labelled left or right by its cluster, and the routing threshold $\tau$
on $t_1$ is set to the midpoint of the two cluster centroids' $t_1$
coordinates. A near-vertical separation -- one in which the two
centroids barely differ in $t_1$ -- is rejected, because the
prediction-time routing rule of Section~\ref{sec:tbtree} of the main
paper uses $t_1$ alone. The split is impurity-invariant and is selected
by a separate constructor flag.

The motivation is that a cut on $t_1$ alone cannot separate regimes
whose signature lives in the inner relation $u_1$ rather than in $t_1$;
clustering the $(t_1,u_1)$ pairs can recover such a separation. Across
the three datasets studied in this paper, however, the $k$-means split
was never the strongest configuration, which is why it is documented
here rather than in the main text.

\paragraph{Simulation.}
On the four-regime simulation, the $k$-means split is impurity-invariant
by construction and lands between the variance and slope criteria on
both predictive $R^2$ and regime-$F_1$; the Scenario~A figures are the
last two rows of Table~\ref{tab:scenA-headline} below.

\paragraph{Archaeological glass.}
On the archaeological-glass calibration of \citet{Lemberge2000glass},
the $k$-means split recovers the known glass typology less cleanly than
the exhaustive $t_1$ cut. On the reference-oxide scope (Na$_2$O \&
Fe$_2$O$_3$) its centroid separation along $t_1$ is too weak to clear
the rejection threshold, so it returns a single leaf
($\text{ARI} = 0$, no split). On the all-oxide scope it grows four
leaves with $\text{ARI} = 0.432$, comparable to $\Phi^u$ at $0.453$ but
with higher test RMSEP on the major oxides (Na$_2$O $1.764$ vs.\
$\Phi^u$'s $1.646$, SiO$_2$ $2.422$ vs.\ $1.596$, CaO $2.092$ vs.\
$0.905$).

\paragraph{Energy efficiency.}
On the energy-efficiency benchmark of Section~\ref{sec:energy} of the
main paper, the $k$-means split is the most parsimonious \tbtree\
configuration -- six leaves and $108$ stored coefficients -- at a mean
test $R^2$ of $0.947$ (heating $0.961$, cooling $0.933$), a little
below the $0.960$ of the $\Phi^u$ tree but at fewer leaves.

% ---------------------------------------------------------------------
\subsection{Cross-dataset summary}\label{sec:impurity-supp-summary}
% ---------------------------------------------------------------------

Table~\ref{tab:impurity-summary} consolidates the comparison.
$\Phi^u$ is the headline criterion on all three benchmarks; the
supplementary criteria each win on a specific configuration --
slope-contrast on regime-$F_1$ for the simulation, kurtosis on PbO
RMSEP for glass, $k$-means on parameter parsimony for energy
efficiency. Each is therefore worth exposing as a constructor flag,
but none displaces $\Phi^u$ as the multi-purpose default.

\begin{table}[htb]
\centering
\caption{Cross-dataset comparison of the four impurity criteria. The
headline metric for each benchmark is in bold; bold-italic marks the
configuration where the row's criterion strictly beats $\Phi^u$.}
\label{tab:impurity-summary}
\begin{tabular}{lccc}
\toprule
Criterion & Sim.\ B ($R^2$ / $F_1$) & Glass B (ARI / PbO) & Energy (mean $R^2$ / leaves) \\
\midrule
$\Phi^u$ (main text)   & $\bm{0.953}$ / $0.679$ & $\bm{0.453}$ / $0.315$ & $\bm{0.960}$ / $10$ \\
Slope-contrast         & $0.984$ / $\bm{\mathit{0.745}}$ & --   / -- & $0.708$ / $10$ \\
Kurtosis-based         & $0.892$ / $0.963$ & $0.413$ / $\bm{\mathit{0.262}}$ & $0.955$ / $6$ \\
$k$-means $(t_1, u_1)$ & $0.938$ / $0.986$ & $0.432$ / $0.367$ & $0.947$ / $\bm{\mathit{6}}$ \\
\bottomrule
\end{tabular}
\end{table}

% =====================================================================
\section{Simulation Scenario A: slope-only regime mixture}\label{sec:scenA}
% =====================================================================

Scenario A is the four-regime mixture of
Section~\ref{sec:sim-design} of the main paper with regime mean
shift $y_{\text{scale}} = 0$, i.e.\ all four regimes share a common
response mean of zero and differ only in their slope structure
$\bP_k,\bQ_k$. This scenario is the harder of the two for any
splitting rule that relies on within-leaf response homogeneity, since
the marginal $\bY$-distribution is the same in every regime; only
the conditional distribution $\bY \mid \bX$ varies. We run the
scenario at three noise levels $\sigma_\varepsilon \in \{0.1,0.3,0.5\}$
and report averages over five seeds.

\subsection{Predictive accuracy and regime recovery}

Table~\ref{tab:scenA-headline} reports test $R^2$ (averaged across
responses) and the regime-$F_1$ score under a leaf-to-regime majority
assignment, for the three impurity criteria with and without the CV
pruning gate. The variance impurity dominates as a predictor across
all three noise levels, with mean test $R^2$ of $0.694$, comparable
to the slope impurity's $0.667$. The slope impurity, on the other
hand, produces the better leaf-to-regime mapping, with mean
regime-$F_1$ of $0.965$ at $\gamma = 0$ against the variance
impurity's $0.959$. The $k$-means split is impurity-invariant by
construction and falls between the two.

\begin{table}[htb]
\centering
\caption{Scenario A (slope-only mixture): test $R^2$ averaged across
responses and regime-$F_1$ score for the three impurity criteria,
averaged over five seeds and three noise levels. Tree depth fixed at
three; $n_{\min} = 15$.}
\label{tab:scenA-headline}
\begin{tabular}{lcc}
\toprule
Configuration & mean $R^2$ & regime $F_1$ \\
\midrule
\tbtree\ variance, $\tau_{\text{cv}}=-1$  & $0.694$ & $0.959$ \\
\tbtree\ variance, $\tau_{\text{cv}}=0.01$ & $0.697$ & $0.961$ \\
\tbtree\ slope, $\gamma=0$, $\tau_{\text{cv}}=-1$  & $0.667$ & $0.965$ \\
\tbtree\ slope, $\gamma=0$, $\tau_{\text{cv}}=0.01$ & $0.668$ & $0.963$ \\
\tbtree\ $k$-means, $\tau_{\text{cv}}=-1$  & $0.679$ & $0.954$ \\
\tbtree\ $k$-means, $\tau_{\text{cv}}=0.01$ & $0.681$ & $0.956$ \\
\bottomrule
\end{tabular}
\end{table}

\subsection{Why variance impurity is the better predictor here}

This may seem counter-intuitive: Scenario A's $\bY$-distribution is
constant across regimes, so a within-leaf $\bY$-variance criterion
should not by itself capture the regime boundary. The mechanism is
that the variance impurity is computed on the leaves induced by a cut
on $t_1$, and conditioning on $t_1 < \tau$ implicitly conditions on
the regime mix in each child even when $\mathbb{E}[\bY \mid k] = 0$
marginally. The variance impurity is therefore not strictly response-
marginal; it inherits the regime separation from the score-plane cut
and amplifies it through the loading orthogonality between regimes.

\subsection{Why slope impurity is the better leaf-to-regime mapper}

The slope impurity scores a cut by the within-child OLS residual on
the parent's score matrix. When all four regimes share zero mean,
this residual is precisely the regime-specific slope \rev{mismatch} -- a
direct measure of how cleanly the cut separates regimes. The variance
impurity, by contrast, can be lured into balanced cuts even when they
mix two regimes whose slope structure cancels out in the response
mean.

\subsection{Noise sensitivity}

At the lowest noise level $\sigma_\varepsilon = 0.1$, all three
impurities reach mean $R^2$ above $0.85$ and regime-$F_1$ above
$0.98$. At $\sigma_\varepsilon = 0.5$ the per-regime signal is weak,
mean $R^2$ falls to $0.35$--$0.45$ and regime-$F_1$ to the $0.45$--$0.65$
range. The slope impurity at $\tau_{\text{cv}} = 0.01$ degrades fastest
at this noise level, dropping to a 3-leaf tree on some seeds; the
variance impurity remains more robust.

% =====================================================================
\rev{\section{Mixture proportions and sample size}\label{sec:pi-n-supp}

Section~3.3 of the main paper characterizes the population behaviour
of the coskewness objectives under a two-regime mean-shift mixture:
the raw criterion is maximized exactly at the regime-shift direction
$\bm{\delta}_x$, the score-standardized criterion at the Fisher
discriminant direction
$(\bm{\Sigma}^x_{\mathrm{w}})^{-1}\bm{\delta}_x$, the signal carries
the prefactor $\pi(1-\pi)(2\pi-1)$, vanishing at $\pi=\tfrac12$,
and the empirical criterion pays sixth-order-moment sampling noise of
order $n^{-1/2}$. This section traces those statements empirically
across a grid of mixture proportions
$\pi \in \{0.5, 0.55, 0.65, 0.75, 0.9\}$ and sample sizes
$n \in \{100, 250, 500, 1000, 2500, 5000, 10000\}$, with $50$
replicates per cell.

\paragraph{Design.}
Each replicate draws regime labels with proportion $\pi$ and, within
each regime, $\bX \in \R^{n\times 6}$ Gaussian with a common
within-regime covariance carrying correlation $0.95$ between $x_1$
and $x_2$ and unit variances; the regime mean shift is
$\bm{\delta}_x = 1.5\,\bm{e}_1$. The response block
($q=2$) follows a common within-regime linear relation loading only
on $x_4$. This is a genuine second-order signal that acts as a decoy for
the covariance objective, orthogonal to the shift, plus a regime
shift $\bm{\delta}_y = (1, 0.4)^\top$ and Gaussian noise. Because the
within-regime correlation between $x_1$ and $x_2$ is strong, the raw target $\bm{\delta}_x$ and the Fisher target
$(\bm{\Sigma}^x_{\mathrm{w}})^{-1}\bm{\delta}_x$ are far apart, so
the two variants are genuinely distinguishable. Three objectives are
compared: covariance, raw coskewness, and standardized coskewness,
each fitted as a single-component decomposition on the centred
(unscaled -- the design is already on a common unit) blocks. Two
metrics are recorded per fit: the absolute cosine between the first
$\bX$-weight and the objective's own population target, and the
adjusted Rand index \citep{HubertArabie1985} between the true regime
labels and a two-means partition of the first score $t_1$. The study
is reproduced by \texttt{simulation\_pi\_n\_study.py} in the
package's \texttt{examples/} folder.

\begin{figure}[htbp]
\centering
\includegraphics[width=0.96\linewidth]{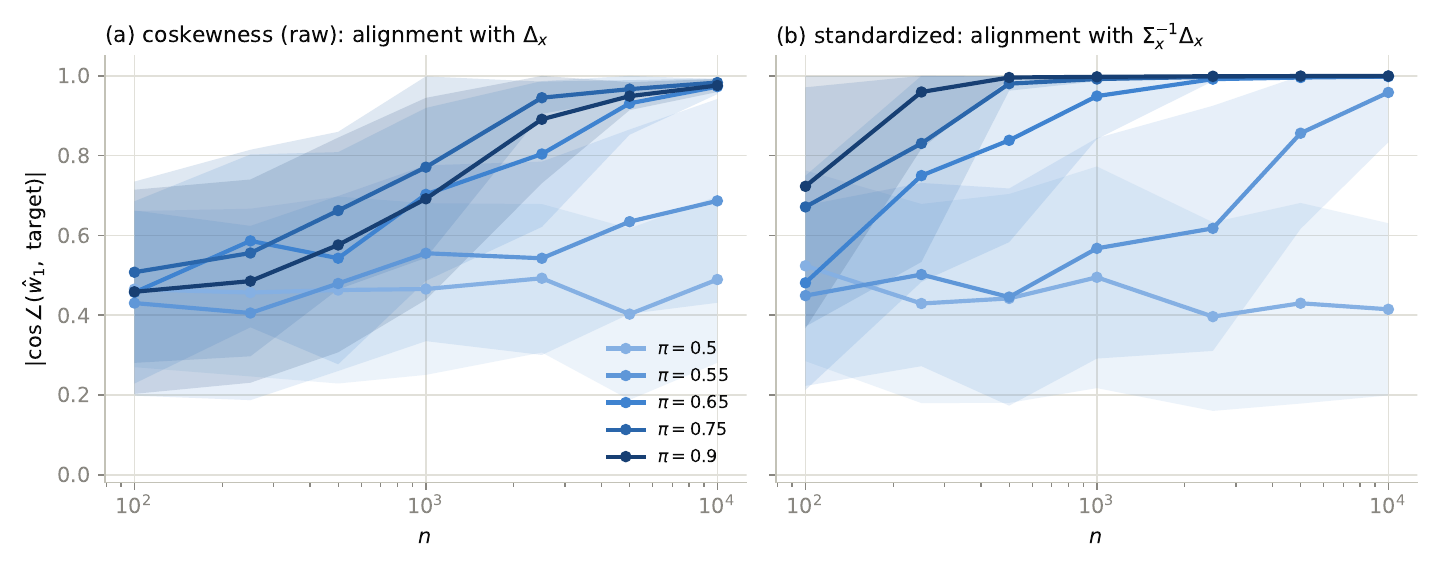}
\caption{Direction recovery across mixture proportions and sample
sizes: mean absolute cosine between the estimated first weight and
the population target of Proposition~\ref{prop:coskew-ident} (panel a, raw coskewness) and
of its corollary (panel b, standardized), $\pm 1$ standard deviation
over $50$ replicates.}
\label{fig:pi-n-recovery}
\end{figure}

\begin{figure}[htbp]
\centering
\includegraphics[width=0.96\linewidth]{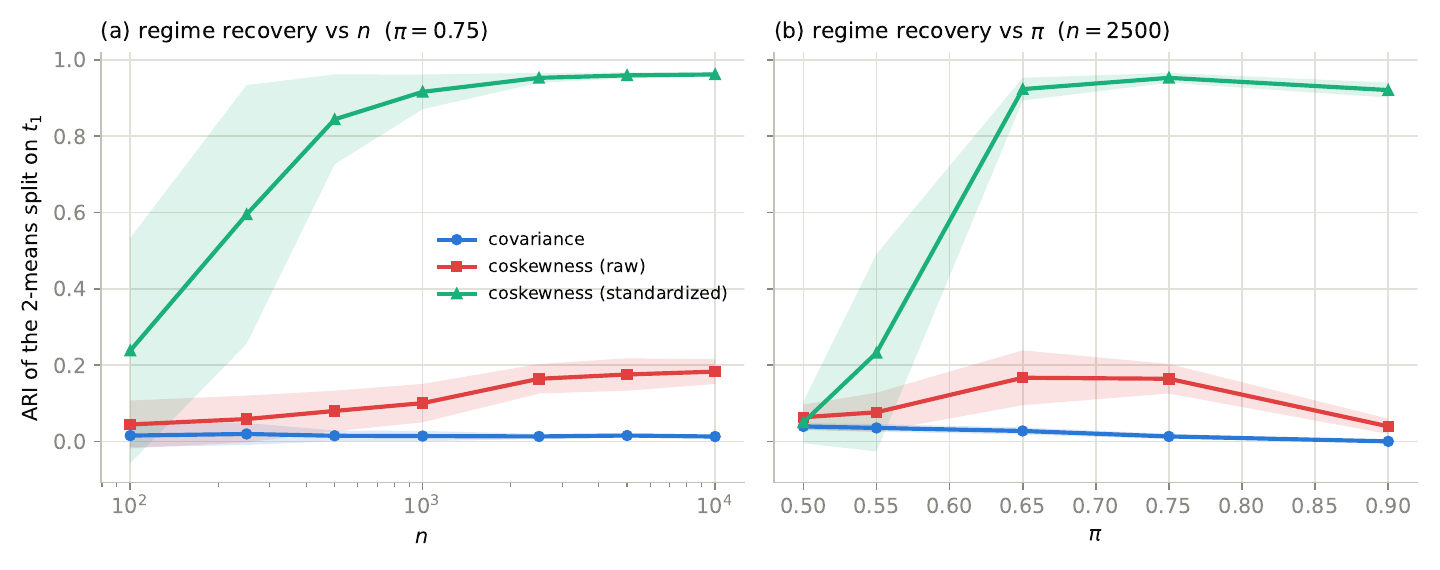}
\caption{Regime recovery (ARI of the two-means split on $t_1$) as a
function of the sample size at $\pi=0.75$ (panel a) and of the
mixture proportion at $n=2500$ (panel b), $\pm 1$ standard deviation
over $50$ replicates.}
\label{fig:pi-n-ari}
\end{figure}

\paragraph{Findings.}
Figure~\ref{fig:pi-n-recovery} confirms the three population
statements. First, at $\pi = 0.5$ both third-moment variants stay
flat far from their targets at every $n$: the balanced mixture
carries no third-order signal, and no sample size repairs that. Second,
the onset of consistency moves with the imbalance exactly as the
$(2\pi-1)^{-2}$ scaling of the main paper's Remark~\ref{rem:samplesize} predicts: at
$\pi=0.55$ the raw criterion only reaches mean alignment $0.69$ at
$n=10^4$, while at $\pi = 0.75$ it passes $0.95$ by $n=2500$ and at
$\pi=0.9$ by $n=5000$. Third, the standardized variant converges
markedly faster: at $\pi=0.75$ it is essentially exact
(alignment $0.98$) from $n=500$ onwards, because whitening removes
the high-variance $x_1$--$x_2$ ridge that otherwise competes with
the shift signal in finite samples. The covariance objective, by
contrast, is not sample-size limited at all but target-limited: it
converges to its own population maximizer, which the within-regime
$x_4$ relation dominates, so its alignment with the shift direction
stays near chance level ($\approx 0.3$) at every $n$.

Figure~\ref{fig:pi-n-ari} translates direction recovery into regime
recovery. The standardized variant dominates: at $\pi = 0.75$ its
ARI climbs from $0.24$ at $n=100$ to $0.95$ by $n=2500$. The raw
variant plateaus near $0.18$ even once its direction estimate is
exact, because the projection of the two regimes onto
$\bm{\delta}_x$ itself overlaps heavily under the strong
within-regime correlation, aligned with the signal-to-noise argument
for the Fisher direction in the main paper's Corollary~\ref{cor:coskew-std}. The
covariance objective recovers nothing (ARI $\le 0.04$) at any
sample size. Panel (b) shows the transition in $\pi$: at $\pi=0.5$
all three objectives fail, by $\pi = 0.65$ the standardized variant
is already near its ceiling, and the mild decline of the raw variant
towards $\pi=0.9$ reflects the shrinking minority regime rather than
a weakening criterion.

\paragraph{Sample-size guidance.}
Read column-wise, the study is the sample-size analysis requested for
the small-$n$ regime: at $n \approx 100$ the third-moment objectives
hold no advantage over the covariance-based state of the art even
under a favourable mixture, consistent with the concrete-slump
finding of Section~\ref{sec:slump-cv} ($78$ training rows), while
from a few hundred rows onwards they transition to
near-perfect regime recovery that the covariance objective never
attains at any $n$.  More precisily, one observes the advantage of the third-moment objectives earlier for the standardized
variant and later for near-balanced mixtures. One caveat delimits the standardized variant's
dominance: this design deliberately satisfies the common
within-regime covariance assumption of the main paper's
Corollary~\ref{cor:coskew-std}. When each regime carries its own
covariance structure -- as in the four-regime generator of
Section~\ref{sec:tbtree-sim} of the main paper, whose regimes have
regime-specific loading matrices -- the pooled whitening misweights
directions and the raw criterion is the preferable choice; see the
guidance in Section~\ref{sec:coskew-theory} and the empirical
comparison across the benchmarks in Sections~\ref{sec:tbtree-sim}
and~\ref{sec:examples} of the main paper.
}

\rev{\section{Sensitivity to the combined-objective weight
$\gamma$}\label{sec:gamma-supp}

Remark~\ref{rem:gamma} of the main paper argues that sensitivity to
the combined-objective weight $\gamma$ is of \emph{selection} type:
the maximizing direction is piecewise (nearly) constant in $\gamma$,
jumping once between the covariance and the coskewness basin, so any
metric of a tree grown from it should be plateau-shaped in $\gamma$.
This section traces those plateaus on all three examples of the main
paper. For each benchmark the recommended decoupled tree, with combined
split and covariance based leaves, is fit for $\gamma \in \{0, 0.1, \dots, 1\}$ with the exact settings of the
corresponding main-paper experiment ($\gamma=0$ is exactly the
covariance tree and $\gamma=1$ exactly the pure coskewness split, so each curve spans the full continuum). The simulation panels averagefive seeds per cell; the study is reproduced by
\texttt{gamma\_sweep\_study.py} in the package's \texttt{examples/}
folder.

\begin{figure}[htbp]
\centering
\includegraphics[width=0.96\linewidth]{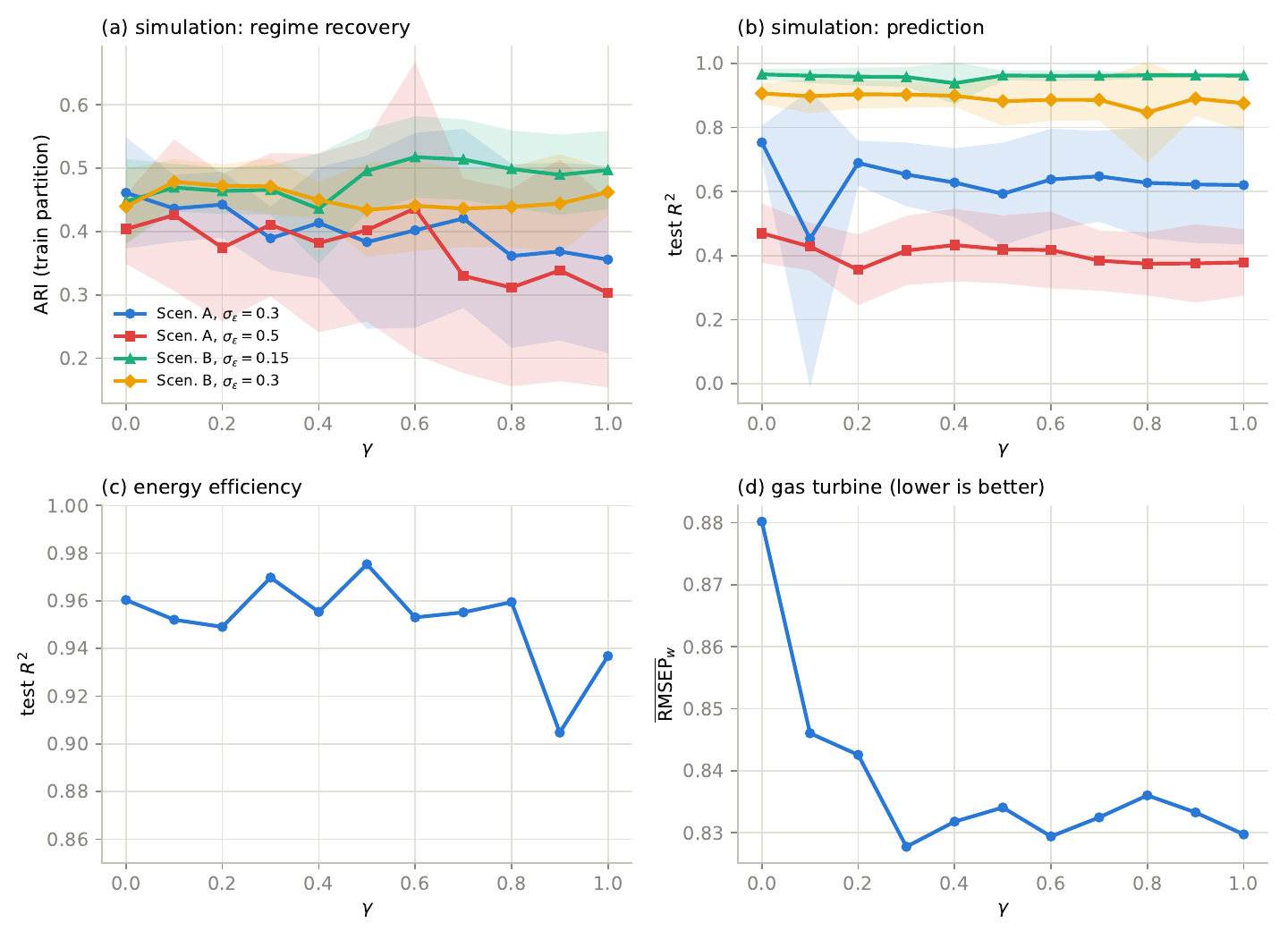}
\caption{Downstream performance of the decoupled combined-split tree
as a function of $\gamma$. (a, b) Four-regime simulation, Scenarios A
and B at two noise levels ($\pm1$ standard deviation over five
seeds). (c) Energy-efficiency test $R^2$. (d) Gas-turbine
scale-weighted RMSEP.}
\label{fig:gamma-sweep}
\end{figure}

Figure~\ref{fig:gamma-sweep} shows the predicted structure. On the
gas turbine (panel d) the covariance tree at $\gamma=0$ scores
$\overline{\mathrm{RMSEP}}_w = 0.880$ and \emph{every} non-zero
$\gamma$ lands in a flat $0.828$--$0.846$ band: the entire benefit of
the third-moment split is captured as soon as the coskewness term is
switched on, and the exact value of $\gamma$ is immaterial. The
energy benchmark (panel c) fluctuates within a two-point band
($R^2 = 0.949$--$0.975$) for $\gamma \le 0.8$, with the maximum at
the default $\gamma=0.5$; the wiggles reflect discrete
re-arrangements of the CV-pruned tree rather than a continuous trend,
and the single dip at $\gamma=0.9$ is recovered by $\gamma=1$. In the
simulation, Scenario B is flat in both metrics across the whole
range, while Scenario A at $\sigma_\varepsilon=0.3$ displays the
two-plateau structure most cleanly: a covariance plateau at
$\gamma=0$ ($R^2=0.75$) and a coskewness plateau for
$\gamma \ge 0.2$ ($R^2 \approx 0.62$--$0.65$), joined at
$\gamma = 0.1$ by the one genuinely unstable point of the whole
study. There, the across-seed standard deviation of $R^2$ spikes to
$0.46$, because $\gamma=0.1$ sits at the crossing weight
$\gamma^\ast$ of Remark~\ref{rem:gamma}, where the global maximizer
is decided by seed-level noise. That single boundary point is the
empirical signature of the basin transition; on either side of it,
performance is stable in $\gamma$.

The practical reading is the one given in
Section~\ref{sec:coskew-obj} of the main paper: cross-validate
$\gamma$ over the coarse grid $\{0, 0.25, 0.5, 0.75, 1\}$,  shince fine
grids resolve nothing that the plateaus do not already show, and
fall back on the default $\gamma = 0.5$ when no tuning budget is
available, which on these benchmarks is never far from the best achievable value.
}

\rev{\section{Sensitivity to the CV-gate threshold
$\tau_{\text{cv}}$}\label{sec:cvtol-supp}

The cross-validation termination gate of
Section~\ref{sec:cv-gate} of the main paper accepts a split only if
it reduces the parent's variance-normalized $K$-fold CV error by at
least a relative $\tau_{\text{cv}}$, with default
$\tau_{\text{cv}} = 0.01$. Table~\ref{tab:cvtol-supp} sweeps the
threshold on the energy-efficiency benchmark for the paper's
recommended configuration (decoupled combined split with
$\gamma=0.5$, covariance leaves, $K=5$, all other settings as in the
main paper's Section~\ref{sec:energy});
$\tau_{\text{cv}} = -1$ disables the gate entirely.

\begin{table}[htb]
\centering
\caption{CV-gate threshold sweep on the energy-efficiency benchmark:
mean test $R^2$, number of leaves and depth of the fitted tree.}
\label{tab:cvtol-supp}
\begin{tabular}{lccc}
\toprule
$\tau_{\text{cv}}$ & mean test $R^2$ & \# leaves & depth \\
\midrule
$-1$ (gate off) & $0.975$ & $11$ & $4$ \\
$0$             & $0.975$ & $11$ & $4$ \\
$0.005$         & $0.975$ & $11$ & $4$ \\
$0.01$ (default)& $0.975$ & $11$ & $4$ \\
$0.02$          & $0.975$ & $11$ & $4$ \\
$0.05$          & $0.975$ & $11$ & $4$ \\
$0.10$          & $0.961$ & $5$  & $4$ \\
\bottomrule
\end{tabular}
\end{table}

The fitted tree is byte-identical from gate-off up to
$\tau_{\text{cv}} = 0.05$: on this benchmark every split the greedy
recursion commits reduces the normalized CV error by more than five
percent, so the gate's role at the default is that of a safeguard
against marginal splits rather than an active pruner. Only at
$\tau_{\text{cv}} = 0.1$ does the gate start refusing splits,
trading $1.4$ points of $R^2$ for a tree half the size,  a
reasonable exchange when parsimony is at a premium, and a
demonstration that $\tau_{\text{cv}}$ operates as a coarse
model-size dial. The default $0.01$ sits well inside the broad
stable region.
}

\rev{\section{Benchmark protocol, uncertainty and computational
cost}\label{sec:protocol-supp}

This section documents the full protocol behind the real-world
comparisons of Section~\ref{sec:examples} of the main paper and adds
uncertainty quantification through repeated splits.

\paragraph{Software and hardware.}
All experiments run under Python~3.9.6 with NumPy~1.26.4,
SciPy~1.13.1, scikit-learn~1.6.1 and pandas~2.3.3; the temporal
convolutional network uses PyTorch~2.2.2 (CPU) and the Cubist
reference the \texttt{cubist} package~1.0.0. \tbtree\ and the
twoblock estimators come from the accompanying open-source package.
Timings are single-machine wall-clock times on an
$8$-core Intel Core i9-9880H (2.3\,GHz, 32\,GB RAM); no GPU is used.

\paragraph{Tuning policy.}
No model -- neither \tbtree\ nor any competitor -- was tuned on test
data, and no per-model hyperparameter search was performed. Every
competitor runs at its library defaults or at a widely used reference
capacity: CART at scikit-learn defaults (grown to purity); the random
forest with $200$ trees; the multilayer perceptrons with fixed
architectures ($(64,32)$ on the energy data; $(64,32,16)$ inside a
feature- and target-standardizing pipeline on the gas turbine), ReLU
activations, Adam, and early stopping on a held-out validation
fraction; Cubist with $10$ committees and $5$ nearest neighbours; the
TCN with the standard architecture of \citet{BaiKolterKoltun2018}
($4$ dilated residual blocks of $32$ channels, kernel $3$, dropout
$0.1$, $24$-hour input windows, Adam at learning rate $10^{-3}$,
batch size $256$, early stopping on the last $15\%$ of the training
period). \tbtree\ likewise uses its defaults with the settings stated
in the main text; on the energy data its per-node $5$-fold CV selects
component counts only. All estimators are seeded, and all data
splits are generated from fixed seeds, so every number in the paper
is reproducible from the scripts shipped with the package
(\texttt{energy\_repeated\_splits.py},
\texttt{gt\_rolling\_origin.py} under \texttt{examples/}).

\paragraph{Computational cost.}
Table~\ref{tab:protocol-supp} lists wall-clock fit times: the median
over the twenty energy splits, and the times on the gas-turbine
$2011$--$2014$/$2015$ reference split. The \tbtree\ energy times are
dominated by the per-node cross-validated component selection (dozens
of twoblock fits per node); a single fixed-parameter tree, as on the
gas turbine, fits the $29{,}349$-row training block in $14$--$20$
seconds ($46$ for the standardized variant, which adds one economy
SVD per extraction).

\begin{table}[htb]
\centering
\caption{Wall-clock fit times (seconds): median over the twenty
energy-efficiency splits and single-fit times on the gas-turbine
reference split.}
\label{tab:protocol-supp}
\begin{tabular}{lcc}
\toprule
Model & energy (median s) & gas turbine (s) \\
\midrule
flat \twoblock                             & $0.005$ & $0.1$ \\
\tbtree\ coskew split, cov leaves          & $36$    & $19$ \\
\tbtree\ coskew split (standardized)       & --      & $46$ \\
\tbtree\ cov+coskew split, cov leaves      & $44$    & $14$ \\
CART regression tree                       & $0.007$ & $0.7$ \\
Random forest, $200$ trees                 & $0.7$   & $10$ \\
MLP, ReLU                                  & $2.3$   & $38$ \\
Cubist ($10$ committees)                   & --      & $35$ \\
TCN ($4$ blocks)                           & --      & $151$ \\
\bottomrule
\end{tabular}
\end{table}

\paragraph{Repeated splits: energy efficiency.}
Table~\ref{tab:energy-repeated} repeats the comparison of the main
paper's Table~\ref{tab:energy} over twenty random $80/20$ splits
(seeds $0$--$19$). The combined-split \tbtree\ and the deep CART tree
are statistically tied, with the tree achieving the tie at roughly
two per cent of CART's leaf count; the random forest is one point
ahead; the MLP is the least stable competitor. Every single-split
value quoted in the main text lies within one standard deviation of
its repeated-split mean.

\begin{table}[htb]
\centering
\caption{Energy efficiency over twenty random $80/20$ splits: mean
$\pm$ standard deviation of mean test $R^2$.}
\label{tab:energy-repeated}
\begin{tabular}{lc}
\toprule
Model & mean test $R^2$ \\
\midrule
flat \twoblock                          & $0.897 \pm 0.012$ \\
\tbtree\ coskew split, cov leaves       & $0.922 \pm 0.024$ \\
\tbtree\ cov+coskew split, cov leaves   & $0.971 \pm 0.014$ \\
CART regression tree                    & $0.971 \pm 0.005$ \\
Random forest, $200$ trees              & $0.981 \pm 0.003$ \\
MLP $(64,32)$, ReLU                     & $0.952 \pm 0.024$ \\
\bottomrule
\end{tabular}
\end{table}

\paragraph{Rolling-origin evaluation: gas turbine.}
Random resplits would destroy the temporal-drift design of this
benchmark, so uncertainty is quantified by a rolling-origin
evaluation instead: three temporal folds with test years $2013$,
$2014$ and $2015$, each trained on all preceding years and scored by
the scale-weighted $\overline{\mathrm{RMSEP}}_w$ computed with that
fold's own test-set scale. Table~\ref{tab:gt-rolling} reports the
per-fold values and their mean $\pm$ standard deviation. Two
conclusions stand out. First, the fold effect dominates: $2014$ is
hard for every method and $2013$ easy, with fold-to-fold standard
deviations of $0.11$--$0.26$ that dwarf the between-model gaps within
the leading group. The decoupled \tbtree\ variants, the TCN, the
random forest and the MLP are therefore statistically
indistinguishable on this benchmark; the CART tree is the only method
clearly behind. Second, the score-standardized coskewness tree is
the most stable \tbtree\ variant across folds
($0.80 \pm 0.13$), consistent with its scale-invariant objective.
It is also noted that the TCN entry of the main paper's
Table~\ref{tab:gasturbine} was refreshed by refitting under the
software stack documented above ($\overline{\mathrm{RMSEP}}_w =
0.813$ against $0.840$ in the original submission, identical
architecture and seed): neural sequence models are not
bit-reproducible across deep-learning library versions, a further
reason to read single-split rankings through the uncertainty lens of
Table~\ref{tab:gt-rolling}.

\begin{table}[htb]
\centering
\caption{Gas turbine, rolling-origin evaluation:
$\overline{\mathrm{RMSEP}}_w$ per temporal fold (test year, trained
on all preceding years) and mean $\pm$ standard deviation across
folds.}
\label{tab:gt-rolling}
\begin{tabular}{lcccc}
\toprule
Model & $2013$ & $2014$ & $2015$ & mean $\pm$ sd \\
\midrule
\tbtree\ coskew split, cov leaves      & $0.629$ & $0.990$ & $0.830$ & $0.82 \pm 0.18$ \\
\tbtree\ coskew split (standardized)   & $0.661$ & $0.909$ & $0.829$ & $0.80 \pm 0.13$ \\
\tbtree\ cov+coskew split, cov leaves  & $0.629$ & $0.965$ & $0.834$ & $0.81 \pm 0.17$ \\
TCN ($4$ blocks)                       & $0.667$ & $0.948$ & $0.813$ & $0.81 \pm 0.14$ \\
Cubist ($10$ committees)               & $0.593$ & $1.098$ & $0.863$ & $0.85 \pm 0.25$ \\
Random forest, $200$ trees             & $0.608$ & $0.923$ & $0.876$ & $0.80 \pm 0.17$ \\
MLP $(64,32,16)$, ReLU                 & $0.605$ & $0.864$ & $0.882$ & $0.78 \pm 0.16$ \\
flat \twoblock                         & $0.753$ & $0.909$ & $0.966$ & $0.88 \pm 0.11$ \\
CART regression tree                   & $0.744$ & $1.245$ & $1.104$ & $1.03 \pm 0.26$ \\
\bottomrule
\end{tabular}
\end{table}
}

\section{Per-node CV findings on the concrete-slump dataset}\label{sec:slump-cv}
% =====================================================================

A useful sanity-check for the CV pruning gate is the concrete-slump
test dataset of \citet{YehSlump2007} (103 rows, $p_x = 7$ ingredients,
$p_y = 3$ slump/flow/strength responses, $75/25$ split). With
$\tau_{\text{cv}} = 0.01$, \tbtree\ produces three leaves at depth
two and a mean test $R^2$ of $0.479$, essentially matching a
200-tree random forest (mean $R^2 = 0.495$) and a 3-hidden-layer
MLP ($R^2 = 0.464$), at a compression ratio of approximately $60\times$
versus the random forest and $300\times$ versus the MLP. A single
CART tree over-fits catastrophically on this small dataset, reaching
depth $11$ and $77$ leaves at a negative test $R^2$. This experiment
uses the covariance split objective; the main paper
(Section~\ref{sec:conclusions}) reports that the coskewness \emph{split}
objective, by contrast, degrades both recovery and prediction on this
same small dataset, so the covariance objective remains the right default
at small $n$.

% =====================================================================
\section{Scalability: the scm20d multi-target benchmark}\label{sec:scm20d}
% =====================================================================

The motivation in Section~\ref{sec:intro} of the main paper turns on
\tbtree\ being practicable at scales where mixed-integer optimal trees
are not, and on its being multivariate by construction. We illustrate
both on \textsf{scm20d}, the largest of the standard multi-target
regression benchmarks \citep{SpyromitrosXioufis2016}: $n = 8966$
supply-chain instances, $p = 61$ predictors and $q = 16$ price targets
($20$-day-ahead means). We use a random $70/30$ train/test split and
report the average relative RMSE (aRRMSE, the standard MTR metric) and
the mean $R^2$ across the $16$ targets, for the flat \twoblock\
baseline and \tbtree\ ($d_{\max}=4$, $n_{\min}=300$, exhaustive split,
$\tau_{\text{cv}}=0.01$) under both impurities, at three Y-component
budgets $K$. Timings are single-core, wall-clock.

\begin{table}[htb]
\centering
\caption{\textsf{scm20d} ($n=8966$, $p=61$, $q=16$): mean test $R^2$,
average relative RMSE (aRRMSE, lower is better) and fit time, for the
flat \twoblock\ and depth-$4$ \tbtree\ under the default variance and
the optional variance$_u$ impurity, at three Y-component budgets $K$.}
\label{tab:scm20d}
\begin{tabular}{llccc}
\toprule
$K$ & model & mean $R^2$ & aRRMSE & fit (s) \\
\midrule
$2$ & flat \twoblock\        & $0.436$ & $0.762$ & $0.1$ \\
    & \tbtree\ variance       & $0.567$ & $0.667$ & $17$ \\
    & \tbtree\ variance$_u$   & $0.575$ & $0.662$ & $34$ \\
\midrule
$4$ & flat \twoblock\        & $0.526$ & $0.698$ & $0.1$ \\
    & \tbtree\ variance       & $0.668$ & $0.584$ & $59$ \\
    & \tbtree\ variance$_u$   & $0.672$ & $0.580$ & $54$ \\
\midrule
$8$ & flat \twoblock\        & $0.578$ & $0.659$ & $0.1$ \\
    & \tbtree\ variance       & $0.705$ & $0.551$ & $103$ \\
    & \tbtree\ variance$_u$   & $0.706$ & $0.550$ & $124$ \\
\bottomrule
\end{tabular}
\end{table}

Three points stand out. First, \tbtree\ improves on the flat
\twoblock\ at every budget (e.g.\ mean $R^2$ $0.578 \to 0.706$ at
$K=8$), empirically confirming nonlinear structure that a single
global linear map cannot capture. Second, every fit completes in
seconds to roughly two minutes on a single core -- one to three orders
of magnitude below the $30$~min--$2$~h per-tree budgets reported for
the MIO route \citep{BertsimasDunn2017} -- so the method is usable at a
scale where optimal trees are not. Third, the variance$_u$ advantage
over the default variance criterion shrinks monotonically as $K$ grows
($\Delta R^2 = 0.008,\,0.004,\,0.001$ at $K = 2,4,8$), exactly the
behaviour anticipated in Section~\ref{sec:impurity} of the main paper:
the two criteria coincide once the number of retained $\bY$-components
$K$ reaches the response rank $q$. This reinforces the main-text view of
$\Phi^u$ as an option for the aggressive-reduction regime rather than a
general default. Tuned ensembles and stacked single-target regressors
reach lower aRRMSE on this benchmark; \tbtree's contribution here is a
single, compact, interpretable multivariate model that fits fast and
dominates the linear baseline.

% =====================================================================
\section{Implementation notes}\label{sec:implementation}
% =====================================================================

The \tbtree\ class lives in \texttt{src/twoblock/tbtree.py} of the
\texttt{twoblock} package, alongside a private \texttt{\_TBNode}
dataclass for tree storage. \rev{The module-level helpers
\texttt{\_search\_cut\_*} implement the three cut-search heuristics
(\texttt{exhaustive}, \texttt{polynomial} and
\texttt{kmeans}).} Per-node cross-validation is
handled by \texttt{\_cv\_select\_params}, which returns
$(\text{params}, \text{score})$, where the score is computed under
the per-response variance broadcast denominator described in
Section~\ref{sec:cv-gate} of the main paper. The pruning gate at the call site in
\texttt{\_fit\_node} forwards the parent's per-response variance via
the \texttt{Ys\_var\_override} argument so parent and child scores
share a denominator.

The plotting helpers used in the example notebooks live in
\texttt{src/twoblock/plots.py}: \texttt{tbtree\_dendrogram},
\texttt{tbtree\_score\_scatter} and
\texttt{tbtree\_leaf\_pred\_vs\_obs}, all returning Plotly figures.

% =====================================================================
% arXiv build: standard author-year style (apalike, ships with TeX Live).
% Single reference list serving both the main text and the supplement.
\bibliographystyle{apalike}
\bibliography{bibliography}
% =====================================================================

\end{document}